\documentclass[12 pt]{article}
\usepackage[font={small}]{caption}
\usepackage[margin=0.5in]{caption}
\usepackage[margin=1in]{geometry}
\usepackage[utf8]{inputenc}
\usepackage{setspace} 
\usepackage{natbib}
\usepackage{graphicx}
\usepackage{blindtext, rotating}
\usepackage{array}
\usepackage{gensymb}
\usepackage{tabularx}
\usepackage{url}
\usepackage{appendix}
\usepackage{amsmath}
\usepackage{authblk}
\usepackage{indentfirst}
\usepackage{makecell}
\title{QRIS: A Quantitative Reflectance Imaging System for the Pristine Sample of Asteroid Bennu}

\author[1]{Ruby E. Fulford}
\author[1]{Dathon R Golish}
\author[1]{Dante S. Lauretta}
\author[1]{Daniella N. DellaGiustina}
\author[1]{Steve Meyer}

\author[2]{Nicole Lunning}
\author[2]{Christopher Snead}
\author[2]{Kevin Righter}
\author[3]{Jason P. Dworkin}
\author[1]{Carina A. Bennett}
\author[1,4,5]{Harold C. Connolly, Jr.}
\author[6]{Taylor Johnson}
\author[1]{Anjani T. Polit}
\author[1]{Pierre Haenecour}
\author[1]{Andrew J. Ryan}

\affil[1]{Lunar and Planetary Laboratory, University of Arizona, Tucson, AZ, USA}
\affil[2]{Astromaterials Research and Exploration Science Division, NASA Johnson Space Center, Houston, TX, USA}
\affil[3]{Solar System Division, NASA Goddard Space Flight Center, Greenbelt, MD, USA}
\affil[4]{Department of Geology, Rowan University, Glassboro, NJ, USA}
\affil[5]{Department of Earth and Planetary Science, American Museum of Natural History, New York, NY, USA}
\affil[6]{Dust Data Management, Tucson, AZ, USA}

\date{Submitted 28 February 2024}

\begin{document}

\maketitle
\begin{abstract}
   The Quantitative Reflectance Imaging System (QRIS) is a laboratory-based spectral imaging system constructed to image the sample of asteroid Bennu delivered to Earth by the Origins, Spectral Interpretation, Resource Identification, and Security\textendash Regolith Explorer (OSIRIS\textendash REx) spacecraft. The system was installed in the OSIRIS\textendash REx cleanroom at NASA's Johnson Space Center to collect data during preliminary examination of the Bennu sample. QRIS uses a 12-bit machine vision camera to measure reflectance over wavelength bands spanning the near ultraviolet to the near infrared. Raw data are processed by a calibration pipeline that generates a series of monochromatic, high-dynamic-range reflectance images, as well as band ratio maps, band depth maps, and 3-channel color images. The purpose of these spectral reflectance data is to help characterize lithologies in the sample and compare them to lithologies observed on Bennu by the OSIRIS\textendash REx spacecraft. This initial assessment of lithological diversity was intended to help select the subsamples that will be used to address mission science questions about the early solar system and the origins of life and to provide important context for the selection of representative subsamples for preservation and distribution to international partners. When QRIS imaged the Bennu sample, unexpected calibration issues arose that had not been evident at imaging rehearsals and negatively impacted the quality of QRIS data. These issues were caused by stray light within the lens and reflections off the glovebox window and interior, and were exacerbated by the sample's extremely low reflectance. QRIS data were useful for confirming conclusions drawn from other data, but reflectance and spectral data from QRIS alone unfortunately have limited utility. 
\end{abstract}
\section{Introduction}

 NASA's New Frontiers 3 mission, OSIRIS\textendash REx (Origins, Spectral Interpretation, Resource Identification, and Security\textendash Regolith Explorer), explored the near-Earth asteroid (101955) Bennu and collected a sample of regolith for delivery to Earth \citep{DanteBook}. Bennu has a relatively carbon-rich surface dominated by hydrated clay-bearing minerals, magnetite, carbonates, and organics \citep{Vicky2019, dante2019, 2020simon, Kaplan2020, kaplan2021}. The presence of organics on Bennu's surface makes it an object of astrobiological interest. This 0.5 km rocky body is composed of primitive material that may contain clues about early solar system conditions, the delivery of water and organics to Earth, and the origin of life \citep{DanteBook}. 
 
\begin{figure}[h!]
     \centering
     \includegraphics[scale=0.5]{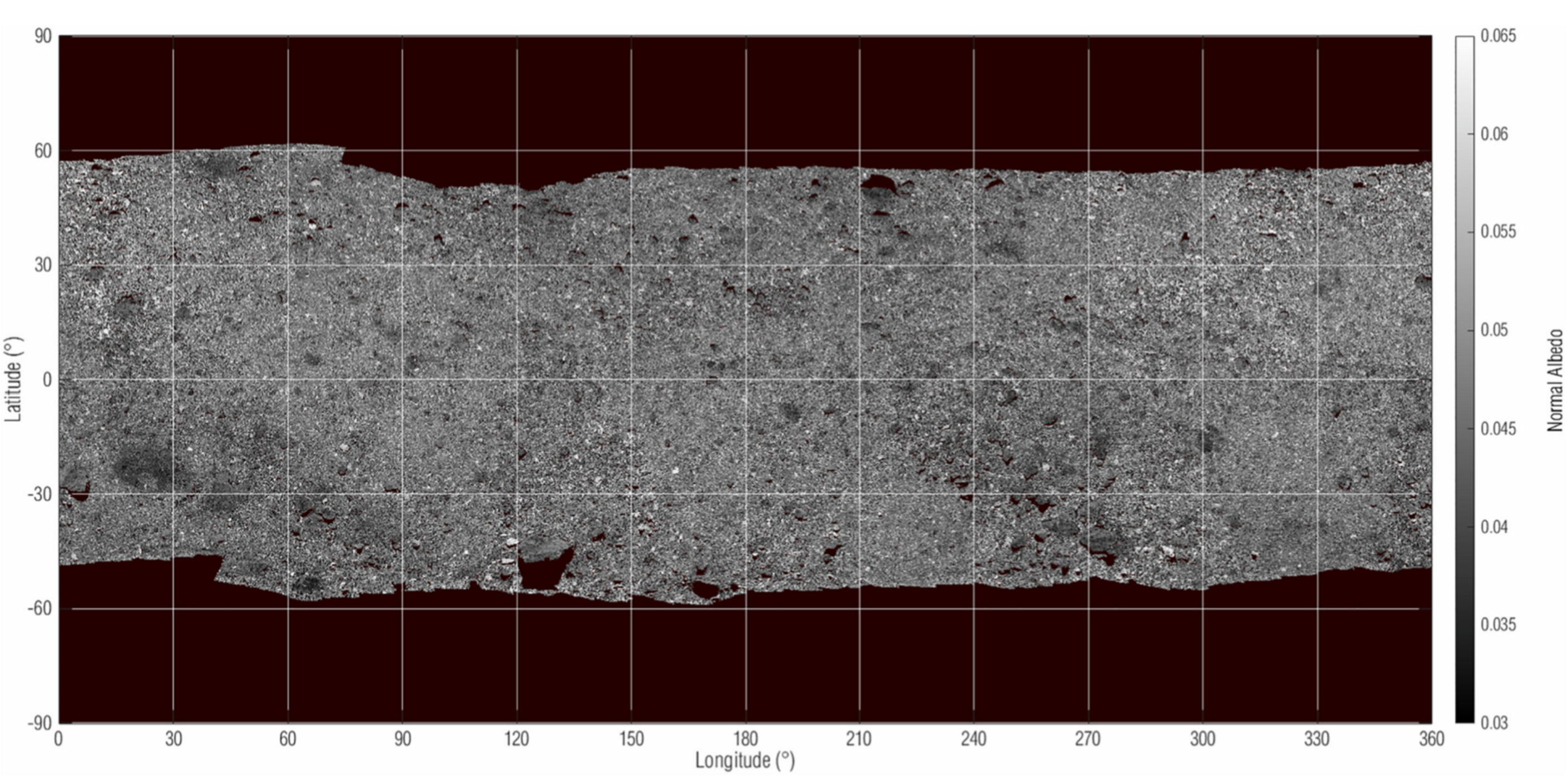}
     \caption{An equirectangular projection of Bennu's normal albedo, reproduced from \cite{golish_map}. The mosaic has a pixel scale of 6.25 cm/pixel and extends to latitudes of about 50$\degree$ north and south.}
     \label{fig:golmap}
 \end{figure}

 The OSIRIS\textendash REx Camera Suite (OCAMS) \citep{dathoncal, rizk} collected extensive information on Bennu's reflectance and color. Data acquired by PolyCam, an OCAMS imager with a broadband panchromatic filter, were used to generate a near-global albedo map in units of reflectance \citep{golish_map}, shown in Fig. \ref{fig:golmap}. This map demonstrates that the majority of Bennu's surface is dark, with normal albedos between 3.0\% and 6.5\%. 

\begin{figure} [h!]
    \centering
    \includegraphics[scale = 0.8]{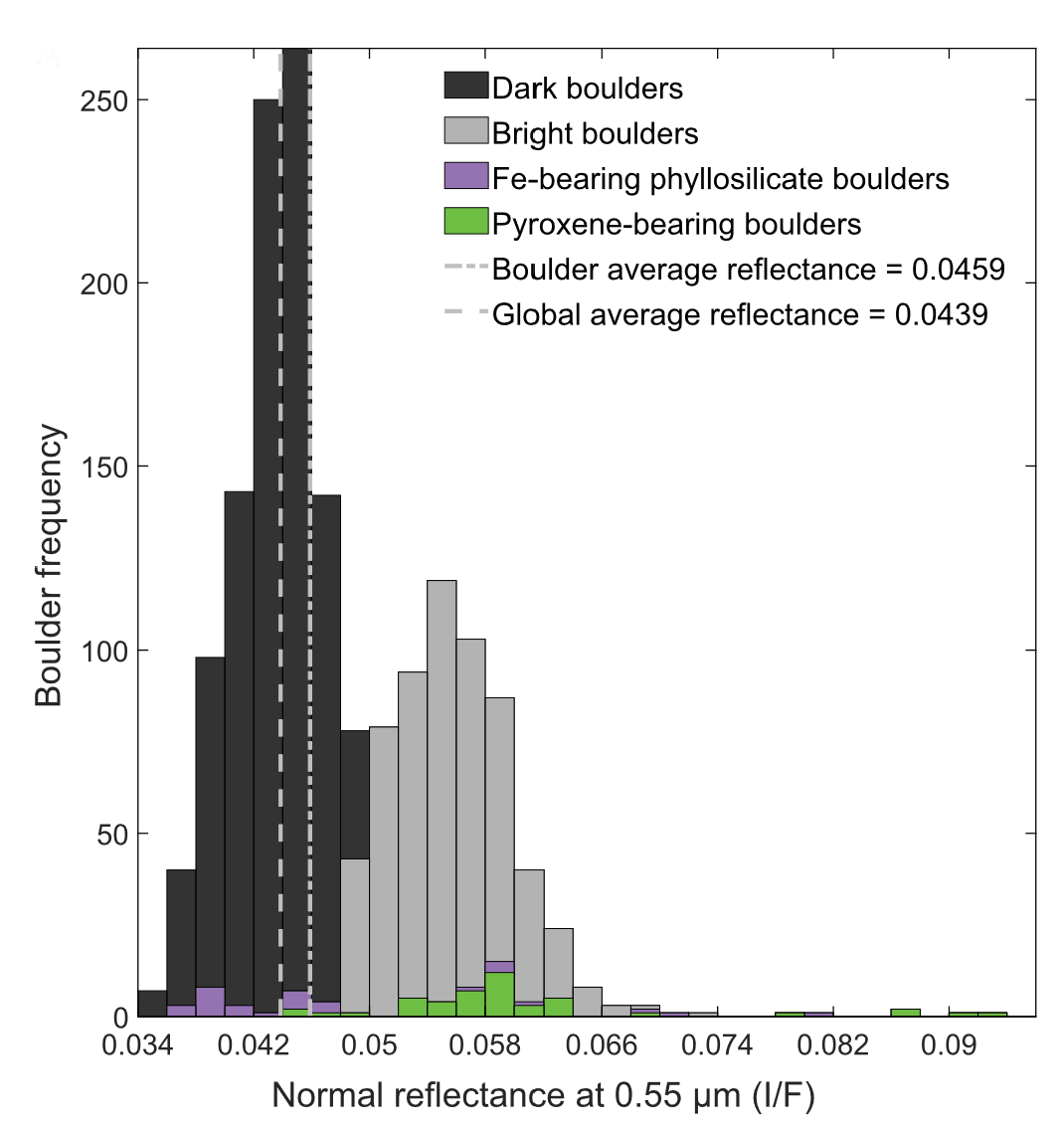}
    \caption{The distribution of normal reflectance values (at 550 nm) of Bennu's surface boulders. Different shades in the histogram indicate the darker boulder population, the brighter boulder population, and minor populations of boulders bearing Fe-phyllosilicate and pyroxene. Vertical lines indicate the average boulder and surface reflectance values. Figure is reprinted from \cite{dellarefl}. Reprinted with permission from AAAS.}
    \label{fig:refl_hist}
\end{figure}

The OCAMS imager MapCam acquired near-global multispectral data in four bands: b$^{\prime}$  (440\textendash 500 nm), v (520\textendash 580 nm), w (670\textendash 730 nm), and x (820\textendash 890 nm). Fig. \ref{fig:refl_hist} shows the distribution of reflectances among Bennu's boulders, as measured by MapCam and described in \cite{dellarefl}. The asteroid's low average reflectance of 4.4 \% may be a result of optically opaque carbon-bearing rocks on its surface. The wide, multimodal reflectance distribution displayed in Fig. \ref{fig:refl_hist} indicates the presence of multiple lithologies. Most of the boulders belong to one of two major lithologies, one very dark and one comparatively bright. The reflectance at 550 nm of the very dark lithology ranges from 3.4\% to 4.9\%, whereas that of the slightly brighter lithology ranges from 4.9\% to 7.4\% \citep{dellarefl}. A few small, bright pyroxene-bearing boulders and clasts, thought to originate from asteroid Vesta, have reflectances up to 26\% \citep{DellaExo}. 

Thermal analysis showed that the two dominant lithologies likely have distinct physical properties \citep{rozitis_2020, rozitis2022},  indicating that they may have been sourced from different depths in Bennu’s parent body. Color variation within individual boulders on Bennu is interpreted to result from varying exposure to space weathering \citep{dellarefl}. 

\begin{figure}[!h]
    \centering
    \includegraphics[scale=0.8]{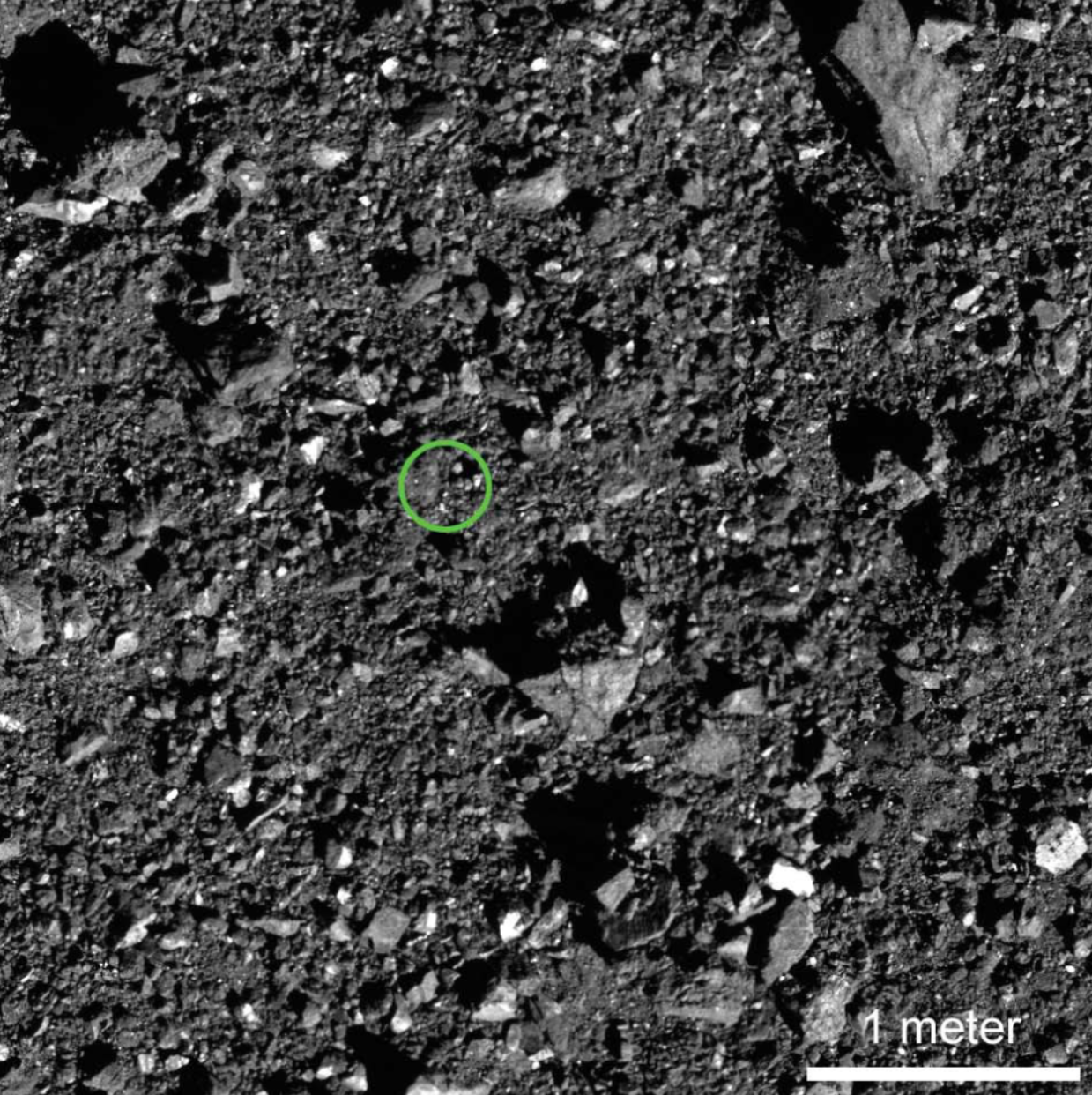}
    \caption{PolyCam image of Hokioi crater, originally published in \cite{Dante2022}. Reprinted with permission from AAAS. The green circle is 32 in diameter and shows where the base of the TAGSAM, the spacecraft's sampling mechanism, contacted the surface.}
    \label{fig:hokoi}
\end{figure}

Despite the distinct reflectances of the two major lithologies, the albedo distribution described by \cite{golish_map} is single-peaked, indicating significant global regolith mixing. The two dominant lithologies are intimately mixed across the surface of Bennu, and both can be found at and near the site in Hokioi crater where OSIRIS-REx collected its sample on October 20, 2020 \citep{rozitis2022, Dante2022}. Therefore, it is probable that both of Bennu's major lithologies are present in the sample. Fig. \ref{fig:hokoi} shows a closeup PolyCam image of the sample collection site. The regolith and boulders display a range of reflectances.

After the OSIRIS\textendash REx sample was delivered to Earth on September 24, 2023, it was transported to curation facilities at NASA's Johnson Space Center (JSC). Preliminary examination of the sample \citep{SAP} focuses on the identification of distinct candidate lithologies to inform allocation of representative subsamples to various teams and institutions involved with the mission. Up to 25 wt\% of the sample will be selected by the OSIRIS\textendash REx science team to address the mission's driving hypotheses, which relate to determining Bennu's composition, its origins and evolution including that of the parent body(ies), and potential contamination introduced during sample collection and recovery \citep{SAP}. 

The Quantitative Reflectance Imaging System (QRIS) is a laboratory-based spectral imaging system, assembled from commercial products, designed for initial color and reflectance imaging of the Bennu sample while it is in the controlled environment of the glovebox in the OSIRIS-REx cleanroom \citep{Righter2023} at JSC. A key motivation behind the development of QRIS is the ability to distinguish between Bennu's two major lithologies, as well as any minor lithologies present. Because both major lithologies are dark, and their peak reflectances differ only by a few percent, QRIS is designed to precisely measure small differences in the reflectance of dark material. 

\begin{figure}
    \centering
    \includegraphics[scale = 0.05]{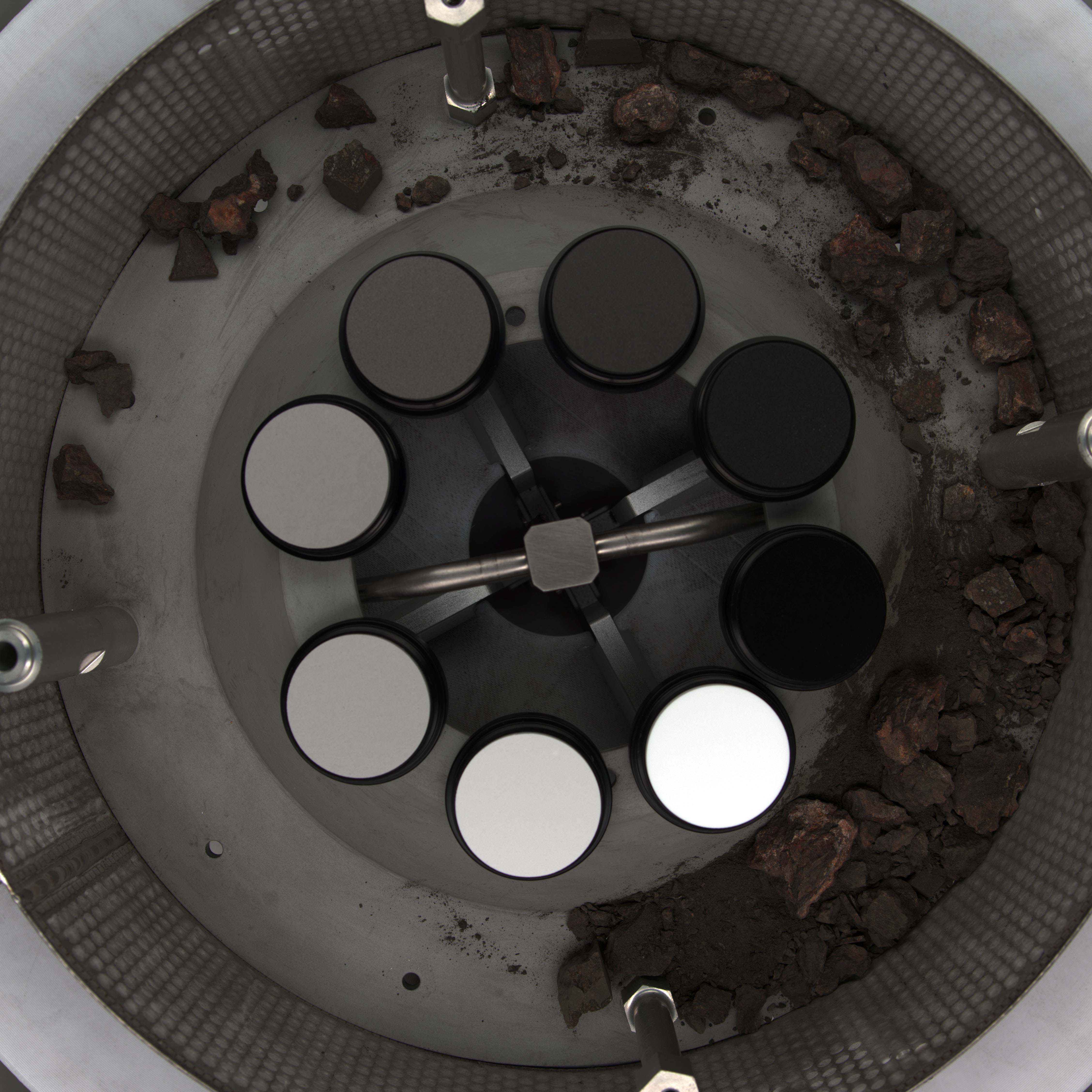}
    \includegraphics[scale=0.05]{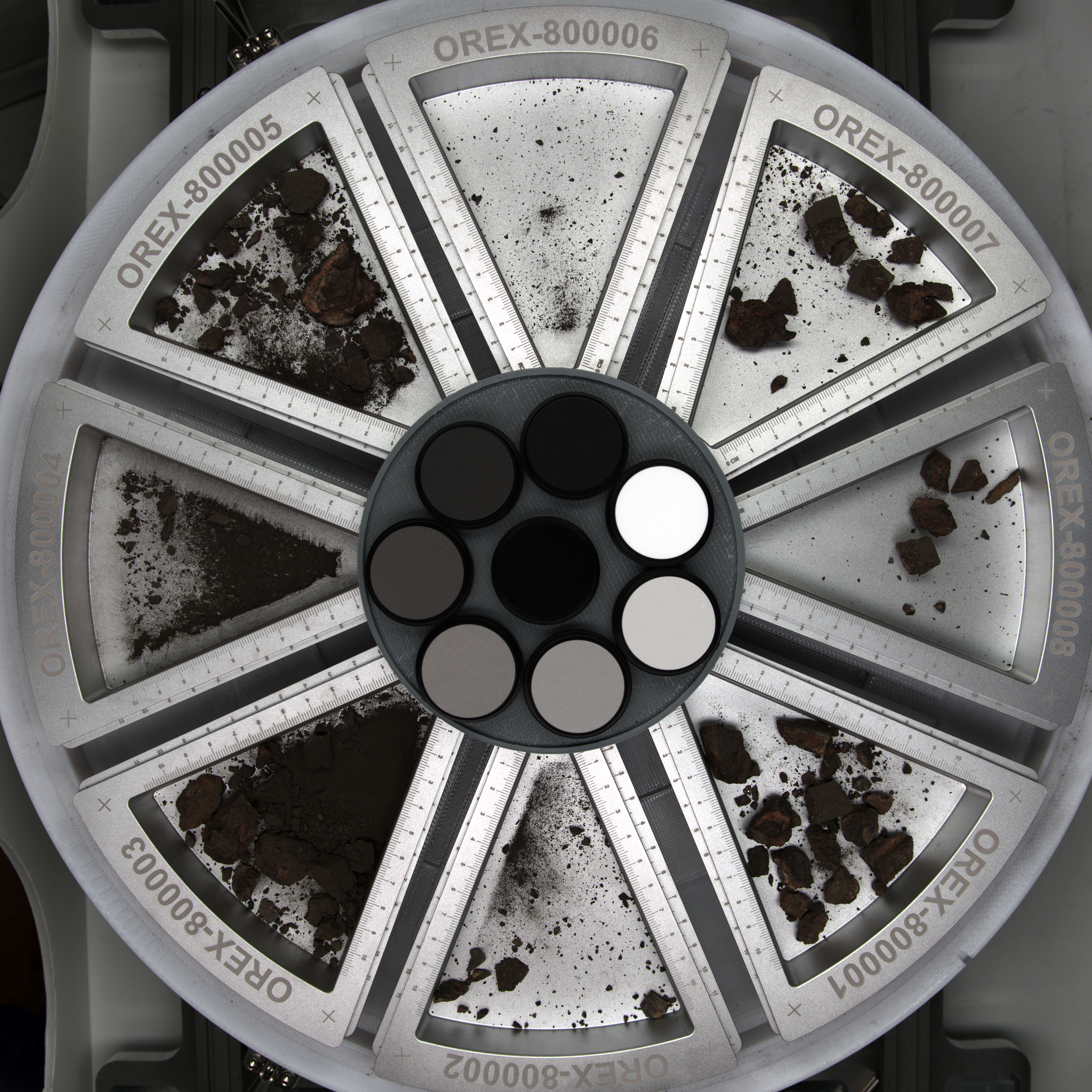}
    \caption{Left: An analog sample in a model of the TAGSAM head. Right: The same analog sample after being poured out of the TAGSAM head into trays. Both RGB images were created by layering images taken by QRIS with red, green, and blue LEDs illuminated during a rehearsal at JSC in March 2023. The eight Spectralon reflectance standards in the field of view, each 3.175 cm in diameter, were used for radiometric calibration for this rehearsal.}
    \label{fig:examples}
\end{figure}

The original plan was for QRIS to image the sample while it was still inside the circular head of the Touch-And-Go Sample Acquisition Mechanism (TAGSAM, \cite{tagsam}). The TAGSAM head dataset provides the science team with a preliminary view of the sample. After TAGSAM head imaging, the JSC curation team would pour the sample into eight wedge-shaped trays to facilitate safe storage and containment. Images of the sample in these trays are the primary QRIS dataset. The science team would use tray images to perform initial analysis of the sample and to characterize potential distinct lithologies on the basis of quantitative reflectance spectra and morphology. QRIS can image regions of interest at higher resolution with a zoom lens. Fig. \ref{fig:examples} shows QRIS images of an analog sample in an engineering model of the TAGSAM head and trays, taken during a rehearsal for initial sample characterization and selection. 

\section{System Design}
To distinguish between Bennu's two dark major lithologies, QRIS must have the ability to detect small differences in reflectance. We aim to measure a reflectance of 2\% within 0.2\%, that is, measure low reflectances with uncertainty $\leq$10\%. QRIS reflectance data must be acquired at multiple wavelength bands so that they can be compared to MapCam data. QRIS utilizes several commercial components (e.g., detector, optics, illumination, diffusing glass, reflectance standards) selected to achieve the aforementioned goals. QRIS is mounted within a custom gantry designed to position the camera and lights above the glovebox window in the OSIRIS\textendash REx cleanroom at JSC. The following subsections explain the system requirements in more detail.

\subsection{Detector}

In the initial phases of QRIS detector selection, we set requirements for its specifications. We require a high-resolution detector to capture millimeter-scale sample features from the detector's position above the glovebox window, $\approx$35 cm above the sample. Our target resolution is 0.1 mm/pixel, corresponding to a pixel count of 4000 pixels in at least one dimension. We determined that a square detector would fit the circular TAGSAM head and trays more efficiently than a rectangular detector. Therefore, we set a minimum resolution requirement of 4000 x 4000 pixels.

Dynamic range (DR) is one of the most important features for the detection of small reflectance variations. Dynamic range for charge-coupled device (CCD) or complementary metal oxide superconductor (CMOS) detectors is defined as the detector’s maximum achievable signal divided by total detector noise \citep{spring_davidson_2006}. Larger dynamic range improves the measurement of fainter regions of a scene, and thus is valuable for observing dark material like Bennu’s regolith.  Dynamic range, often expressed in units of decibels (dB), is calculated using the following equation: 
 \begin{equation}
     DR = 20 \times \log(N_{sat}/N_{noise}),
 \end{equation}
where $N_{sat}$ is the capacity of the detector expressed as a number of electrons, and $N_{noise}$ is the total detector noise expressed as a number of electrons \citep{spring_davidson_2006}.  

A detector needs adequate bit-depth to properly utilize its full dynamic range. One bit corresponds to about 6 dB of dynamic range \citep{spring_davidson_2006}.  A 12-bit detector yields 4096 grayscale levels, meaning every 0.2\% reflectance increment would correspond to about eight levels, enabling us to achieve our objective of measuring 2\% reflectance within 0.2\%. We determined that a bit depth of 12 and a corresponding dynamic range of 72 dB would be sufficient for distinguishing Bennu's lithologies. 

\begin{table}
    \begin{center}
    \begin{tabular}{ |m{15em}| m{15em}|} 
        \hline \textbf{Property} & \textbf{Value} \\ \hline
        Resolution & 4512 x 4512 pixels (20.20 MP)  \\\hline
        Bit Depth &  8/10/12 bits \\
        \hline Dynamic Range & 70.8 dB  \\ \hline 
        Imaging Sensor & Sony IMX541 \\ \hline
        Camera Sensor Format & 1.1" \\ \hline
    \end{tabular}
    
    \caption{Specifications of the Allied Vision Alvium 1800 U-2040m camera}
    \label{tab:al}
    \end{center}

\end{table}

We selected the Allied Vision Alvium 1800 U-2040m detector, a monochromatic machine vision camera with a CMOS detector. Its resolution of 4512 x 4512 pixels and maximum bit-depth of 12 meet system requirements. The dynamic range of 70.8 dB is slightly lower than the requirement, but it can be easily compensated with high-dynamic-range stacking. The camera's features are summarized in Table \ref{tab:al}.

The camera connects to a computer via USB 3, which allows it to be controlled using Allied Vision's Vimba software \citep{alliedvisionVimbaMachine}. We used Vimba's Python Application Programming Interface (API) \citep{VimbaPython} to write camera control programs. 
\subsection{Optics}
 
 QRIS utilizes an HP-series lens from Edmund Optics with a focal length of 12 mm and a diameter of 48 mm. When the lens is positioned above the glovebox window, $\approx$35 cm above the sample, its field of view is $\approx$35 cm wide and captures the entire 32-cm TAGSAM head or all the sample trays. Images taken with this lens at its nominal height have a pixel scale of $\approx$0.08 mm/pixel at their center. Pixel scale and geometric distortion will be further discussed in Section 4.2.

\begin{figure}[h!]
    \centering
    \includegraphics[scale=0.05]{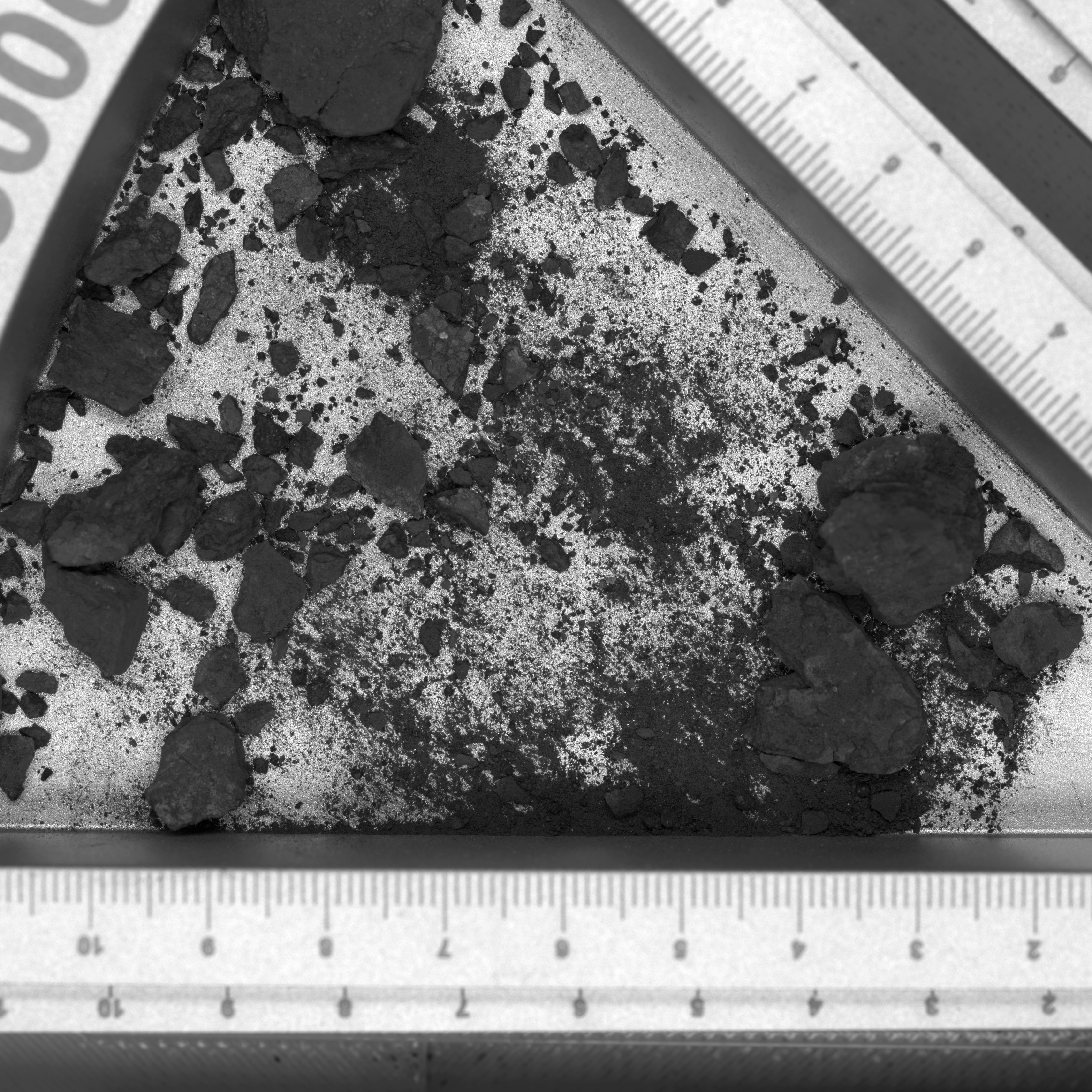}
    \caption{Analog sample in a tray, imaged using QRIS's 50 mm lens. The analog sample pictured is the same as that in Fig. \ref{fig:examples}.}
    \label{fig:zoom_example}
\end{figure}

We also procured an Edmund Optics HP-series lens with a focal length of 50 mm and a diameter of 50 mm for higher-resolution zoom imaging. Images taken with the 50 mm lens have a pixel scale of $\approx$0.02 mm/pixel at their centers. The 50 mm lens will be used to gather detailed images of stones and regions of interest during PE. Fig. \ref{fig:zoom_example} shows an image taken with the 50 mm lens of the same analog sample pictured in Fig. \ref{fig:examples}. 
\subsection{Illumination}
QRIS must image the sample over a range of wavelengths so that reflectance at various wavelengths can be compared. To accomplish this, we illuminate the sample with LED strips covering nine wavelength bands. For broadband white lighting, we use white Waveform Absolute Series LED strip lights. For narrow-band illumination, we use Waveform Simple Color LED strip lights in near ultraviolet (UV), violet, blue, green, amber, red, and near infrared (IR). Our blue, green, and IR LEDs correspond to MapCam's b', v, and x filters, respectively. QRIS will be able to detect a pyroxene absorption feature at 1000 nm via a downturn in the IR band reflectance, just as MapCam detected pyroxene via a downturn in the x band relative to other bands. The peak wavelengths of the blue, green, and IR strips do not exactly match their corresponding MapCam filters because we are limited to wavelengths offered by commercial suppliers. However, QRIS LED band widths and MapCam filter band widths overlap, allowing QRIS to detect spectral features similar to those detected by MapCam. A 700 nm deep red LED strip from LuxaLight, analogous to MapCam's w-band filter, is also included in QRIS specifically to detect the absorption feature at this wavelength by calculating the deep red band depth relative to red and IR reflectance. This absorption feature, indicative of Fe-bearing phyllosilicates, was observed at Bennu via a positive relative band depth in MapCam's w band. 

\begin{figure}[h!]
    \centering
    \includegraphics{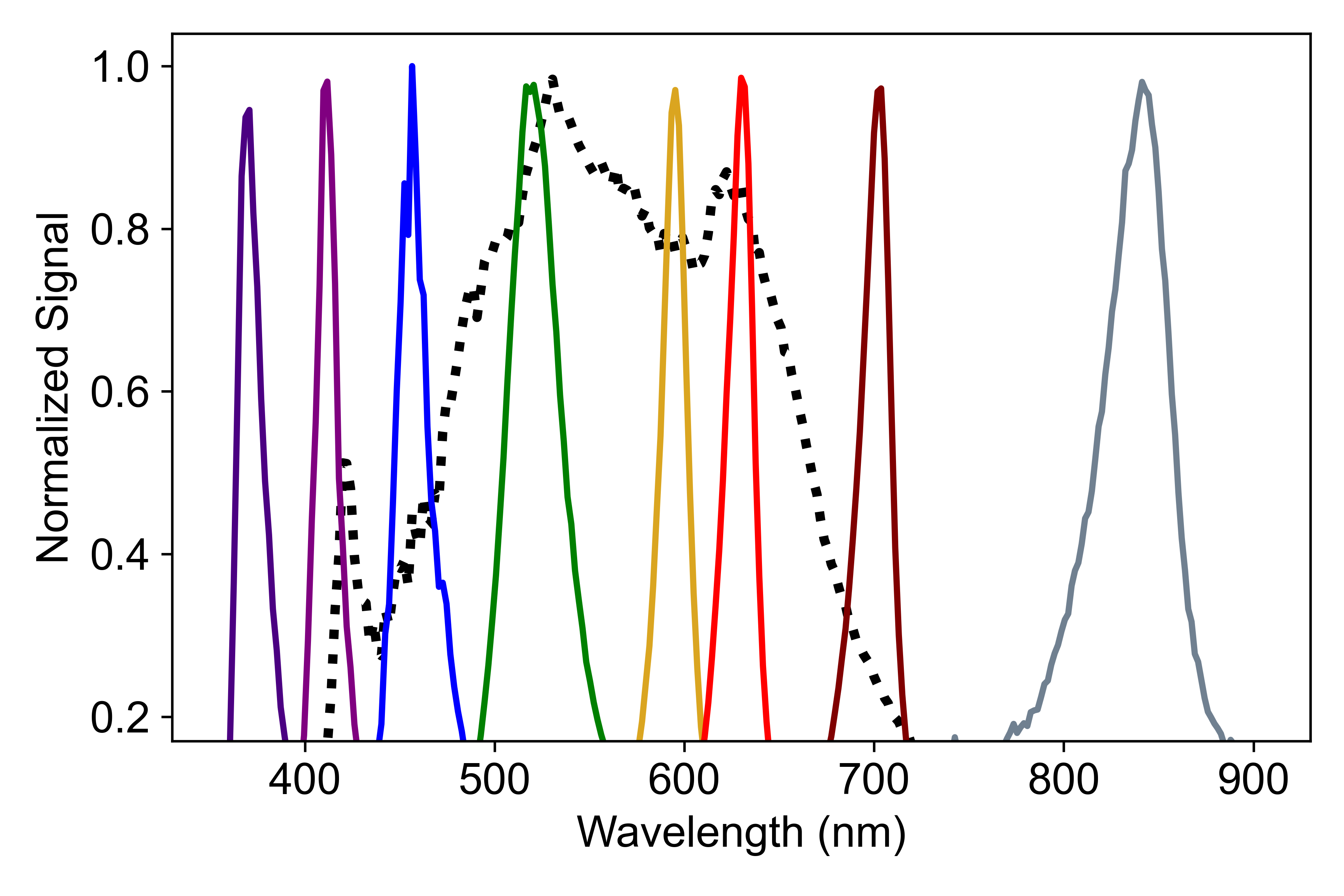}
    \caption{Spectra of the nine LED strips used to illuminate QRIS. Eight of the LED strips emit narrow-band illumination with a distinct peak wavelength. The white LED strip emits a broadband spectrum (black dotted line).}
    \label{fig:ledspectra}
\end{figure}

Peak wavelengths and full width half maximums (FWHMs) for all LED strips, along with corresponding MapCam filters when applicable, are described in Table \ref{tab:lights}. Peak wavelength and FWHM values for all QRIS LED strips are based on spectrometer measurements taken after the LED strips were installed in QRIS. Spectra for all LED strips are displayed in Fig. \ref{fig:ledspectra}.

\begin{table}[h!]
    \begin{center}
   
    \begin{tabular}{|c|c|c|c|c|}
         \hline \textbf{Band} & \textbf{Peak Wavelength (nm)} & \textbf{FWHM (nm)} & \textbf{MapCam Filter} & \textbf{FWHM (nm)} \\ 
         \hline UV & 371& 15.1& &\\
         \hline Violet & 411 & 14.1& & \\
         \hline Blue & 457 & 17.0 & 470 & 30\\
         \hline Green & 521 & 32.6 & 550 & 30\\
         \hline Amber & 594 & 15.5 & & \\
         \hline Red & 629 & 16.9 &&\\
         \hline Deep Red & 700 & 18.11 & 700 & 30\\
         \hline IR & 839 & 38.8 & 855 & 35 \\  
         \hline White & Broadband & N/A & &\\ \hline
    \end{tabular}
    \caption{The peak wavelength and FWHM value for each LED strip used by QRIS, with corresponding MapCam filter information where applicable.}
    \label{tab:lights}
    \end{center}
\end{table}

The LED strips adhere to two aluminum panels positioned on either side of the camera. The panels are enclosed in aluminum boxes that suspend sheets of white diffusing glass 12.7 cm below the lights. The diffusing glass, purchased from Edmund Optics, improves uniformity of the illumination across the field of view. The glass also spatially mixes the colors so that they each illuminate the sample from the approximately same direction. This lessens color artifacts in ratio maps and band depth maps. Each LED panel holds one LED strip of all wavelength bands except UV. The UV LEDs emit less light than the other LEDs and the camera is less sensitive at UV wavelengths, so each panel has two UV LED strips. The two panels can be turned on at the same time or separately. QRIS can therefore image with three different "lighting configurations": both LED panels illuminated, right LED panel illuminated, or left LED panel illuminated. The LED boxes are carefully positioned so as not to reflect in the glovebox window. However, it was not possible to avoid reflections of the light off of the interior of the chamber, due to the extremely shiny metal and the lack off opportunity to test illumination in the glovebox. 

The LED strips are all wired to a digital multiplex (DMX) encoder and Enttec pro interface that connects to a computer via USB, which allows the lights to be controlled remotely via a Python script. The brightness levels of the LED strips are adjustable over an 8-bit range (0 to 255). Brightness levels for imaging sequences are chosen so that images taken with the same exposure time, for all bands except UV, will produce approximately equivalent digital number (DN) signal. Even when set to maximum brightness, the UV LEDs will produce less than half the signal of other bands when imaged with the same exposure.

\subsection{Reflectance Standards}

QRIS's radiometric calibration is established by imaging materials of known reflectance. We use seven Spectralon reflectance standards custom-made by Labsphere, smaller (1.27 cm diameter each) than the eight standards used in rehearsals (Fig. \ref{fig:examples}).  The standards have nominal reflectances of 2\%, 5\%, 10\%, 20\%, 40\%, 75\%, and 99\%. Though we name the standards by these nominal  percentages, the true reflectances of the standards vary with wavelength. Calibrated reflectance values for the standards at each QRIS wavelength band are shown in Table \ref{tab:avg_refls}. We find these reflectance values by measuring the spectrum of each LED band and calculating the average reflectance over the entire normalized spectrum for each band, weighted by intensity. Reflectance spectra for each standard are provided by Labsphere, and have an uncertainty of ~5\%. 

\begin{figure}
    \centering
    \includegraphics{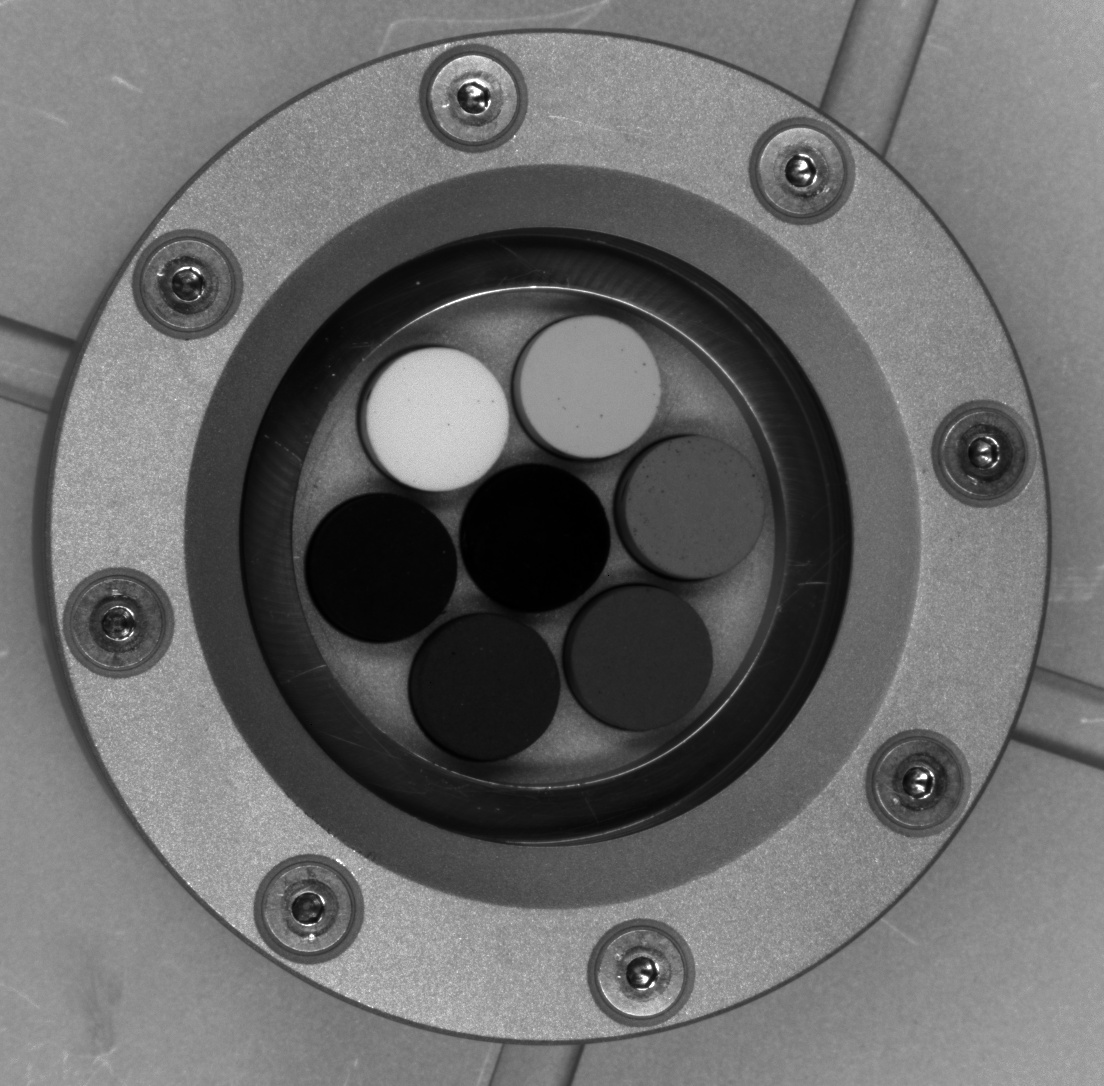}
    \caption{Seven custom Spectralon reflectance standards, each with a diameter of 1.27 cm, enclosed in a gas-tight container behind a sapphire window.}
    \label{fig:standards}
\end{figure}

The reflectance standards are placed in a circular container in the center of the trays holding the Bennu sample. The standards are sealed into the container in a gaseous nitrogen environment. The standards must be imaged alongside the sample in order to be used for radiometric calibration, but they present a contamination risk because they contain carbon. For this reason, the standards' container is gas-tight and the standards are imaged through a 2 mm thick sapphire window. Fig. \ref{fig:standards} shows the standards sealed into this container, imaged by QRIS in the glovebox in the OSIRIS\textendash REx cleanroom at JSC. The sapphire window reduces the apparent reflectance of the standards by about 25\%. Table \ref{tab:sapph} lists the ratios between reflectance standard brightness measured with and without the sapphire window at each wavelength band.

\begin{table}[h!]
    \centering
    \begin{tabular}{|c|c|c|c|c|c|c|c|}
         \hline  & \multicolumn{6}{c}{\textbf{Reflectance of Standard (\%)}} &\\
         \hline \textbf{Nominal} & 2  & 5 & 10 & 20 
         & 40 & 75 & 99
         \\
         \hline White & 1.09 & 4.95 & 10.6 & 20.2 & 38.9 & 73.6 & 97.9 \\
         \hline UV & 1.11 & 4.88 &, 10.5 & 19.5 & 37.5 & 72.2 & 97.6 \\
         \hline Violet & 1.10 &  4.92 & 10.6 & 19.5 & 37.7 & 72.4 & 97.8.\\
         \hline Blue & 1.09 & 4.92 & 10.6 & 19.7 & 38.0 & 72.7 & 97.8 \\
         \hline Green & 1.09 & 4.97 & 10.6 & 20.0 & 38.6 & 73.2& 97.8\\
         \hline Amber & 1.10 & 4.96 & 10.6 & 20.3 & 39.1 & 73.7 &97.8 \\
         \hline Red & 1.10 & 4.97 & 10.7 & 20.5 & 39.4 & 74.0 & 97.8 \\
         \hline Deep Red& 1.12 & 5.01 & 10.7 & 20.7 & 39.7 & 74.3 & 97.7\\
         \hline IR & 1.14 & 5.04 & 10.7 & 20.8 & 39.8 & 74.4 & 97.7\\  
          \hline
    \end{tabular}
    \caption{The nominal reflectances of the Spectralon standards and their actual average reflectances at each QRIS wavelength band.}
    \label{tab:avg_refls}
\end{table}
\begin{table}[h!]
    \centering
    \begin{tabular}{|c|c|c|}
         \hline \textbf{Band} & \textbf{Brightness Ratio} \\ 
         \hline White & 0.741\\
         \hline UV & 0.730 \\
         \hline Violet & 0.730 \\
         \hline Blue & 0.735 \\
         \hline Green & 0.741 \\
         \hline Amber & 0.741 \\
         \hline Red & 0.741 \\
         \hline Deep Red & 0.746\\
         \hline IR & 0.752 \\  
          \hline
    \end{tabular}
    \caption{Ratios of the brightness of the Spectralon reflectance standards when imaged behind a 2 mm sapphire window to their brightness when imaged without the sapphire window. The ratios vary slightly by wavelength.}
    \label{tab:sapph}
\end{table}

\subsection{Software}
Allied Vision provides the Vimba software suite, which can be used to control camera settings and collect images. We write camera and LED control functions using Vimba's Python API \citep{VimbaPython}. All functions that control imaging sequences are based around two key functions: one that triggers the camera to take an image with a user-specified exposure time, and one that turns on an LED strip with a user-specified brightness. Each LED strip corresponds to a channel within the DMX universe. Our script controls the LEDs by setting brightness values for the channels. These basic functions are used to create a variety of more complex functions that trigger image sequences for calibration and data collection. The code utilizes Python packages such as \texttt{Numpy} \citep{numpy}, \texttt{DMXEnttecPro} \citep{dmxenttecpro}, \texttt{open-cv} \citep{opencv_library}, and \texttt{pandas} \citep{pandas}. 

After images are collected, the raw data are run through a \texttt{Python} calibration pipeline that generates 32-bit TIFF images in units of reflectance. The calibration pipeline will be further explained in Section 3.

\subsection{Gantry Design}
\begin{figure}[h!]
    \centering
    \includegraphics[scale=0.055,angle=0]{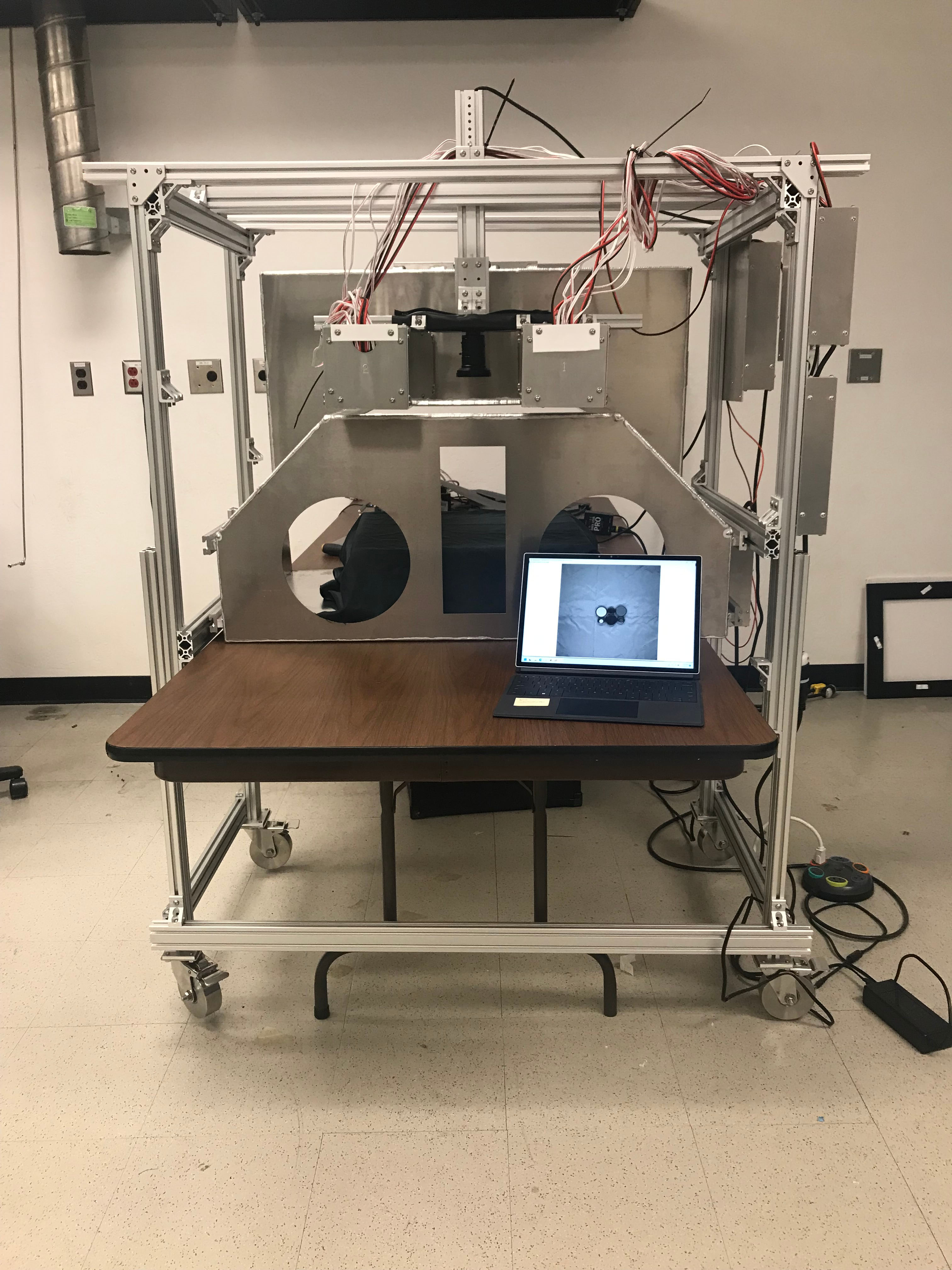}
    \includegraphics[scale=0.055, angle = 0]{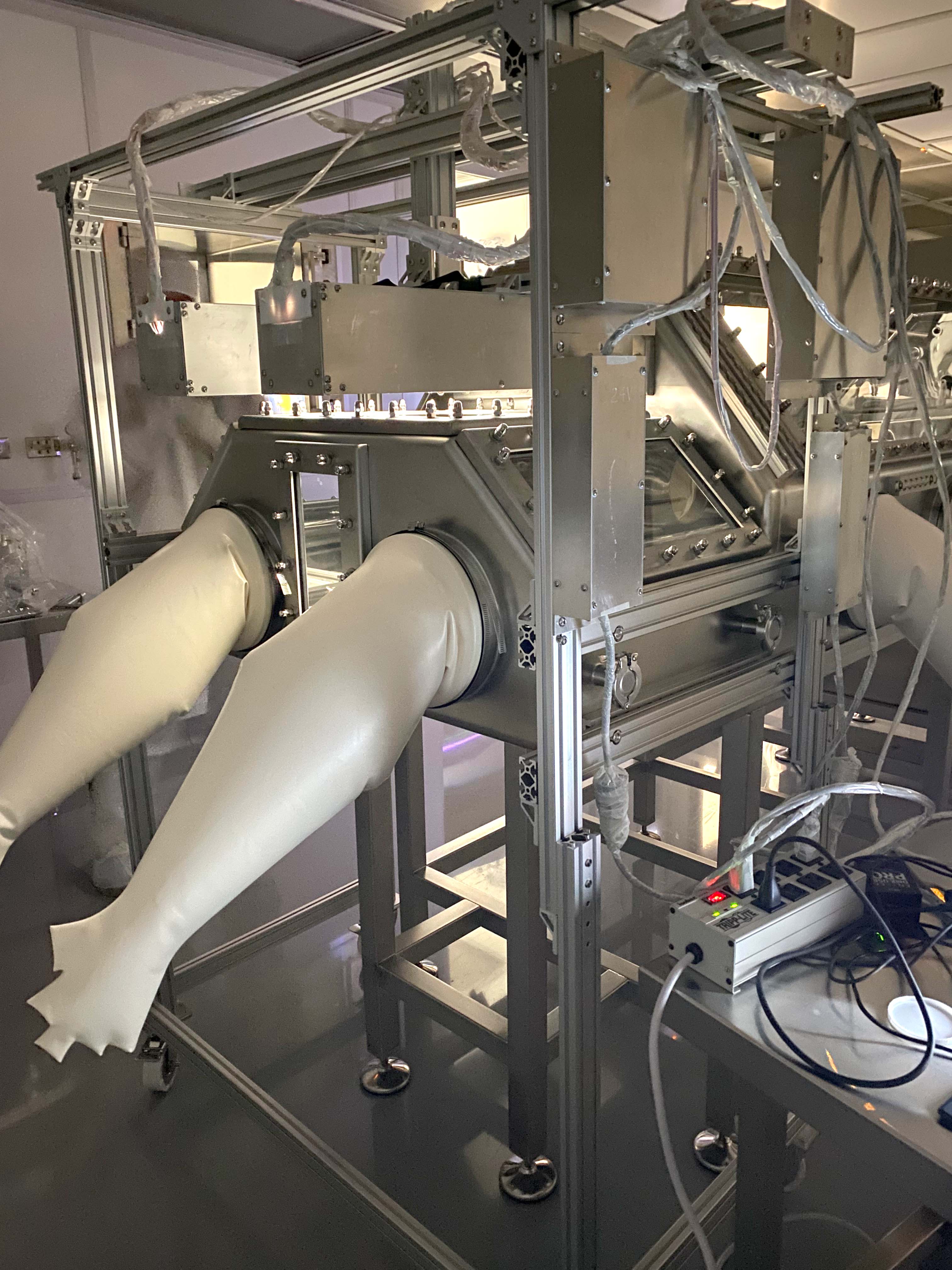}
    \caption{Left: QRIS installed on its gantry over a mock glovebox at the University of Arizona. Right: QRIS installed on its gantry over the real glovebox in the OSIRIS\textendash REx cleanroom at JSC.}
    \label{fig:qris_final}
\end{figure}

The QRIS gantry is designed to suspend the camera and lights over the top window of the glovebox in which the sample is initially examined.  The final version of the QRIS gantry is displayed in Fig. \ref{fig:qris_final}.   

The gantry is largely composed of t-slotted aluminum framing and is designed to allow adjustment of its dimensions and the positions of the camera and lights. Because the glovebox was being manufactured at the same time QRIS was being designed, it was important that the gantry had the flexibility to accommodate changes in the glovebox specifications, such as the width of the mounting points and the height of the sample. As Fig. \ref{fig:qris_final} shows, the gantry has four legs that attach to the glovebox via four US Cargo Control L-tacks on the glovebox's sides. The legs have wheels that allow QRIS to be easily removed from the glovebox once it is detached from the L-tracks. 

The camera and lights are attached to a vertical support beam, which can be raised and lowered to adjust their heights. The camera height can also be changed independently of the lights. Movable components allow us to achieve the desired field of view and eliminate visible reflections off the glovebox window. Even so, the polished stainless steel interior of the glovebox creates complex internal reflections that cannot be fully eliminated by moving the components.

\subsection{Environmental Requirements}
To protect the Bennu sample from contamination and maintain the integrity of the ISO 5 cleanroom and the interior of the nitrogen-purged glovebox, there are restrictions on the materials permitted in these spaces and encapsulation requirements for materials that are out of compliance. A full list of permitted materials can be found in Table \ref{tab:materials} in Appendix A. In addition, no nylon, silicons, or 3D-printed materials were permitted. 

 To comply with the requirements, the gantry and electronics enclosures are made of 6061 and 6063 aluminum. Screws and nuts are made of 316 stainless steel. The camera, LED strips, wiring, and electronics contain non-compliant materials essential to their functions, so these components are sealed inside compliant aluminum enclosures or covered with polytetrafluoroethylene (PTFE, trade name Teflon$^{tm}$) tape, as can be seen in Fig. \ref{fig:qris_final} (right). The lens and the bottom of the LED boxes are left unsealed so that illumination and data collection remain uninhibited. As discussed above, the reflectance standards were encapsulated to permit their use inside the glovebox. The standards and the QRIS flat field (a sheet of white PTFE) were precision-cleaned by the JSC curation team before being placed in the glovebox.

\section{Calibration}

\subsection{Darks}
Dark current is produced by thermally generated electrons and is measured by the detector as signal \citep{janesick}. By design, the detector's on-board electronics estimate and remove dark current before transmitting the data to the user. To verify this, we took several test images with multiple exposure times in a dark environment (no LEDs illuminated, room windows covered). We found that even when using our camera's maximum exposure time of 10 seconds, average dark signal was $\approx8$DN. Dark signal accounts for $<0.5\%$ of signal in long-exposure images taken with the LEDs illuminated, and even less for shorter exposure images. Therefore, we do not correct dark signal in our calibration pipeline. Dark images are nonetheless taken at the start and end of each imaging sequence to verify that the on-board correction remains valid. 

\subsection{Detector Nonlinearity}
\begin{figure}[h!]
    \centering
    \includegraphics[scale=0.35]{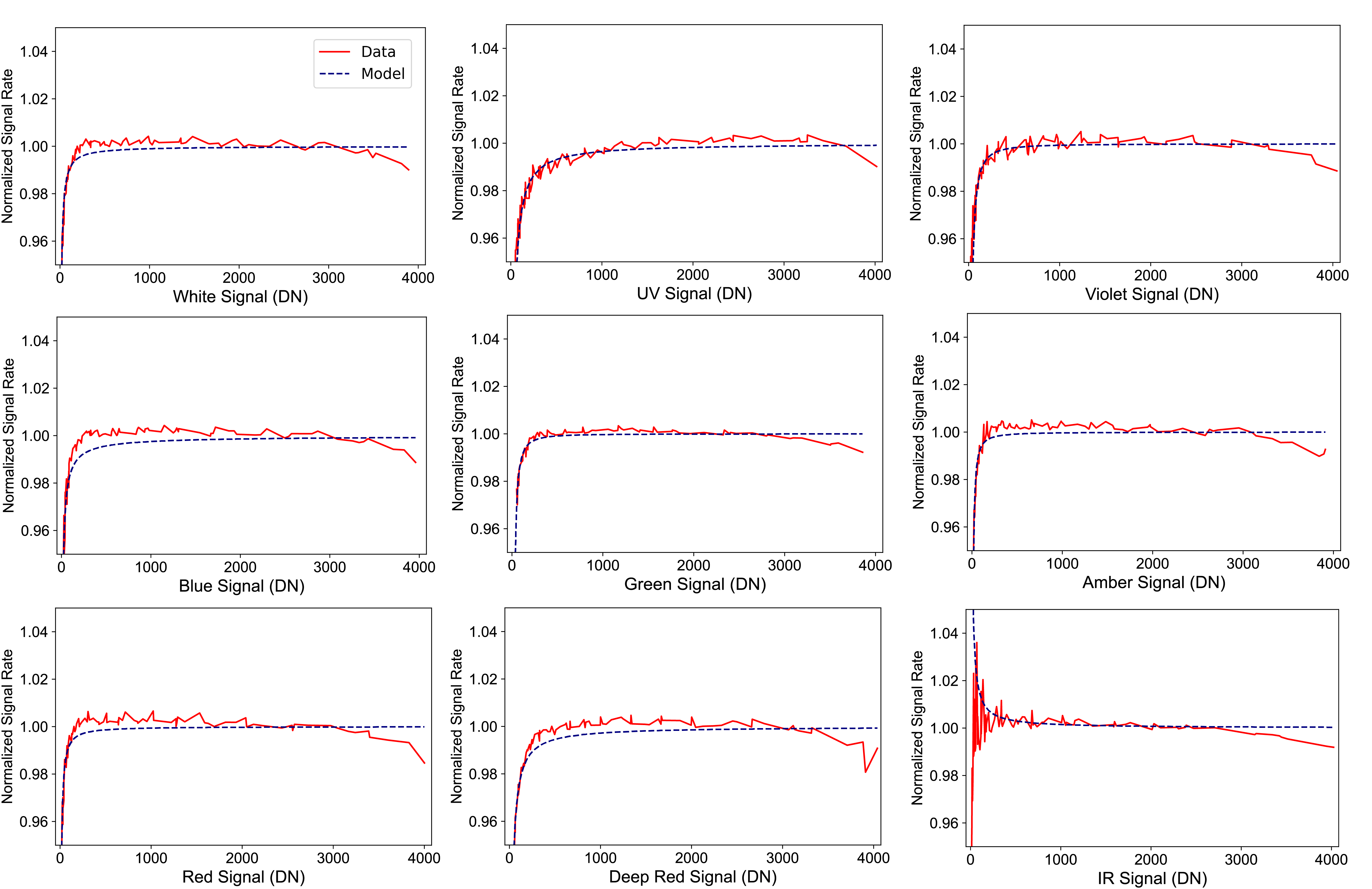}
    \caption{Normalized signal rate vs. signal (red) from imaging of Spectralon reflectance standards under all wavelength bands. The plots display clear deviation from a horizontal line, indicating detector nonlinearity. Models of low-signal nonlinearity are shown in navy. All models are created using Equation \ref{nonlineqn}, with wavelength-dependent parameters described in Table \ref{tab:nonlin}. Plots have been normalized so that average signal rate is 1.}.
    \label{fig:nonlin_models}
\end{figure}

An ideal detector displays a linear relationship between incoming photons and measured signal. Most detectors are nonlinear at very low or very high signal levels \citep{janesick}, typically measuring less signal than should have been detected. To explore the nonlinearity of the QRIS camera, we imaged the Spectralon reflectance standards under each wavelength band at exposure times ranging from 5 to 10,000 ms. The average signal levels of the reflectance standards were extracted from each of these images and used to explore the relationship between normalized signal rate and signal. Signal rate is equal to signal over exposure time. If the detector were perfectly linear, signal rate would be constant across signal level. However, our measured signal rate dips at low and high signals, as shown in Fig. \ref{fig:nonlin_models}. 
We use \texttt{SciPy.curvefit} \citep{SciPy} to model the detector's nonlinearity. \texttt{SciPy.curvefit} is a \texttt{Python} package that utilizes a nonlinear least-squares method to fit a function to the data. We find that low-signal nonlinearity can be modeled by a function of the form 

\begin{equation}
    SR= \frac{m}{S^{n}} + 1 ,
    \label{nonlineqn}
\end{equation}
where \textit{SR} is signal rate, \textit{S} is signal, and \textit{m} and \textit{n} are band-dependent parameters that are calculated by \texttt{SciPy.curve\_fit}. The values of these parameters for each band are reported in Table \ref{tab:nonlin}.

\begin{table}[h!]
    \centering
    \begin{tabular}{|c|c|c|}
         \hline \textbf{Band} & \textbf{m} & \textbf{n}\\ 
         \hline UV & -3.0815 &  0.98129\\
         \hline Violet & -15.655 & 1.4749 \\
         \hline Blue & -0.65332 & 0.80609 \\
         \hline Green &  -14.478& 1.5585 \\
         \hline Amber & -2.2160 & 1.2702 \\
         \hline Red & -1.2369 & 1.0929 \\ 
         \hline Deep Red & -2.3141 & 0.97318 \\
         \hline IR & 1.5000 & 0.99999 \\
         \hline White & -1.0248 & 1.0006 \\
         \hline
    \end{tabular}
    \caption{Nonlinearity parameter values for each wavelength band.}
    \label{tab:nonlin}
\end{table}

\begin{figure}[h!]
    \centering
    \includegraphics[scale=0.35]{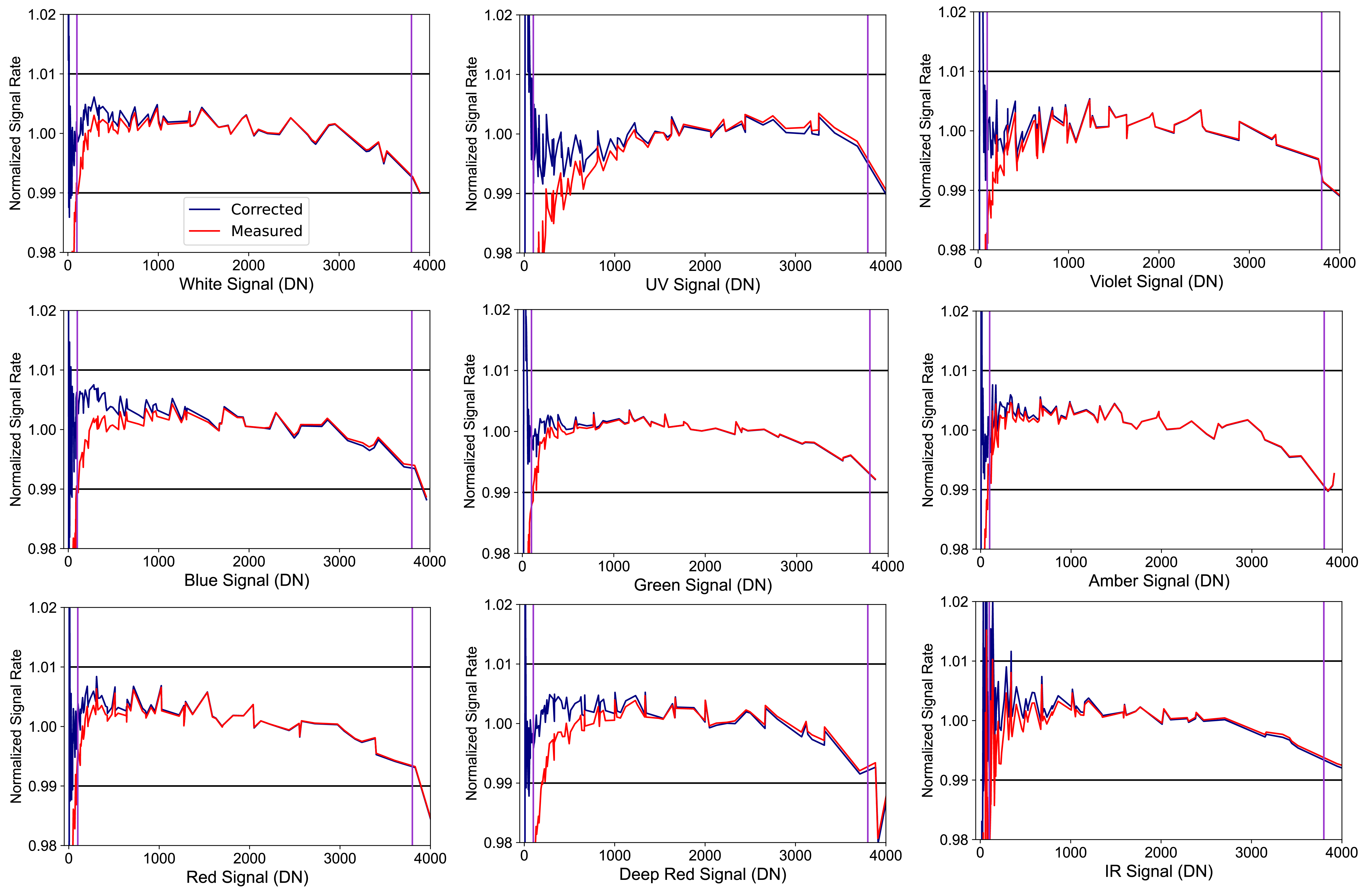}
    \caption{Normalized measured signal rate compared to normalized corrected signal rate for all bands. Each plot has pink vertical lines at 100 DN and 3800 DN to represent the boundaries within which nonlinearity can be corrected. The IR plot has an additional pink line at 200 DN. Black horizontal lines at signal rates of 1.01 and 0.99 show that corrected signal rate does not deviate from 1 by more than 1\% between the pink lines.}
    \label{fig:nonlincorr}
\end{figure}

To correct nonlinearity, each signal measurement in the dataset is divided by 
\begin{equation}
    \frac{m}{{S_{m}}^n}+1,
\end{equation}
where $S_m$ is measured signal. The corrected signal is used to generate a corrected signal rate. Fig. \ref{fig:nonlincorr} shows the corrected signal rates. For all bands except IR, the corrected signal rates do not deviate from linear by more than 1\% within 100 and 3800 DN. Corrected IR signal does not deviate from linear by more than 1\% between 200 and 3800 DN. Corrected signal for all bands does not deviate from linear by more than 2\% above 3800 DN. Nonlinearity correction is applied to pixels with values between 200 and 3800 DN as the first step of the calibration pipeline. 
\subsection{Flat fields}

The position of the LED panels and lens vignetting creates an uneven illumination pattern over the field of view. Uneven illumination must be corrected so that variations in illumination intensity are not misinterpreted as variations in the reflectance of the sample. Non-uniform illumination can be corrected by applying a flat field \citep{janesick}. To create flat fields, we image a matte white PTFE sheet in five slightly different positions with each band and lighting configuration. We median average the five images taken at each band and configuration to minimize the effect of high spatial frequency variations (e.g., small scratches on PTFE) in the flat fields. We further smooth the flats with a circular Gaussian filter to eliminate medium frequency variations. Each flat field is then inverted and normalized so that pixel values are $\geq$ 1. After nonlinearity correction, each image is multiplied by its corresponding flat field to correct brightness variations. We repeat this flat fielding process for the 50 mm zoom lens.

\subsection{High-dynamic-range Stacking}

High-dynamic-range (HDR) images for each wavelength band and LED configuration are created by stacking images taken at multiple exposure times. HDR stacking allows the creation of images that display detailed variation in dark regions of the sample without sacrificing accurate imaging of brighter regions, including the 99\% reflectance standard. In these HDR images, dark regions of the field of view are well-exposed and bright regions remain unsaturated. Exposure times used in the QRIS imaging sequence are 50, 100, 500, 1000, 2000, 5000, and 9999 ms. To combine images of all exposure times, we replace saturated pixels of long-exposure images with corresponding unsaturated pixels of shorter exposure images until no saturated pixels remain. Images are scaled by their relative exposure times (for example, combining a 100 ms image with a 200 ms image requires the signal in the 100 ms image to be multiplied by 2). HDR stacking occurs after flat field corrections are applied. One HDR image is created for each wavelength band and lighting configuration.

\subsection{Radiometric Calibration}
To generate useful reflectance data, HDR images must be converted from units of DN to reflectance. Our seven custom Spectralon reflectance standards are placed in the field of view of sample tray images taken with the 12 mm lens. Our pipeline calculates the average signal over each reflectance standard in every HDR DN image. We use the signal values and reflectance values to create a piecewise reflectance vs signal relationship. For signal values darker than the 2\% target or brighter than the 99\% target, we extrapolate from the nearest piecewise relationship. This function is used to convert the DN signal values of all pixels to reflectance values.

When the 50 mm lens is used, it is not possible for both the sample and the reflectance standards to fit in the field of view. Therefore, we take radiometric calibration image sequences with this lens that only image the reflectance standards, as well as sequences that image the sample. We apply the calibration pipeline to the reflectance standard image sequences and save the average DN signals on each reflectance standard from the HDR images as an array. We then apply the pipeline to the sample images, using the array of average signals to convert DN units to reflectance. It is also not possible to have the reflectance standards in the field of view when imaging the TAGSAM head with the 12 mm lens. The calibration used for the 12 mm images of the sample trays is applied to the 12 mm TAGSAM images.

\subsection{Calibration Pipeline}
Raw images acquired by QRIS are processed by a calibration pipeline written in \texttt{Python} that applies the calibrations described in Sections 3.1\textendash 3.4. Raw, uncalibrated images are referred to as "Level-0" images. These images are 16-bit TIFF files with integer pixel values ranging from 0 to 4095. The first step of the calibration pipeline corrects detector nonlinearity for pixel values between 200 and 3800. The next step of the calibration pipeline applies flat field corrections. Images that have undergone nonlinearity correction and flat field correction are referred to as "Level-1" images. These images are 32-bit floating point TIFF files. The third step of the calibration pipeline creates HDR stacked images for each wavelength band and configuration. These "Level-1 HDR" images are saved as 32-bit floating point TIFF files. The final step of calibration is the radiometric calibration that converts the image to reflectance units. These final, "Level-2" images are 32-bit TIFF files with pixel values between 0 and 1. The uncertainty on the reflectance values is $\leq$10\%, explained further in Appendix B and Table \ref{tab: err}. The full set of narrowband reflectance maps that results from our pipeline can be treated as a spectral data cube for extracting spectra of individual stones or creating band ratios and false color maps.

\section{System Performance}

\subsection{Focus}

Data from rehearsals revealed that the focus of our lenses is wavelength dependent. After extensive focus testing, we determined that using four focus settings for the 12 mm lens produces optimal focus for all wavelengths. UV, violet, and IR images benefit from individual focus settings; all other wavelengths are adequately focused when the camera is focused using amber lighting. During the imaging sequence, the 12 mm lens is refocused under UV, violet, and IR lighting to create "best-focus" images for these wavelengths.  

\begin{figure}[h!]
    \centering
    \includegraphics[scale = 0.5]{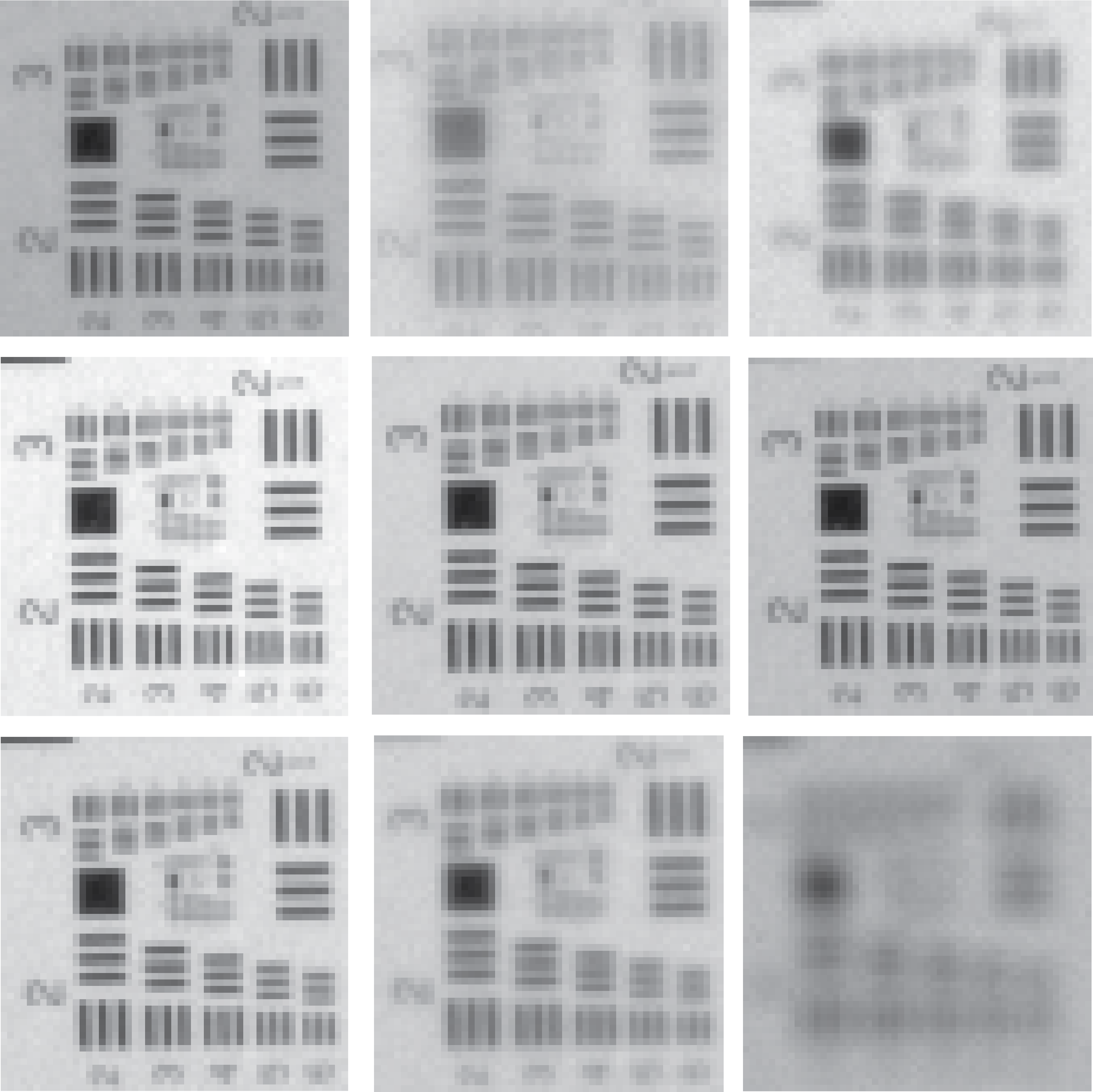}
    \caption{USAF 1951 bar target imaged with the 12 mm lens at all wavelength bands, focused under amber LEDs. Top row, from left to right: white, UV, violet. Center row, from left to right: blue, green, amber. Bottom row, from left to right: red, deep red, IR. The target resolution for the system is represented by Group 2, Element 3 (groups are horizontal, elements are vertical). This element is sufficiently resolved at all wavelength bands when focused under amber illumination, except for violet, UV, and IR.}
    \label{fig:common focus}
\end{figure}

\begin{figure}[ht]
    \centering
    \includegraphics[scale=0.5]{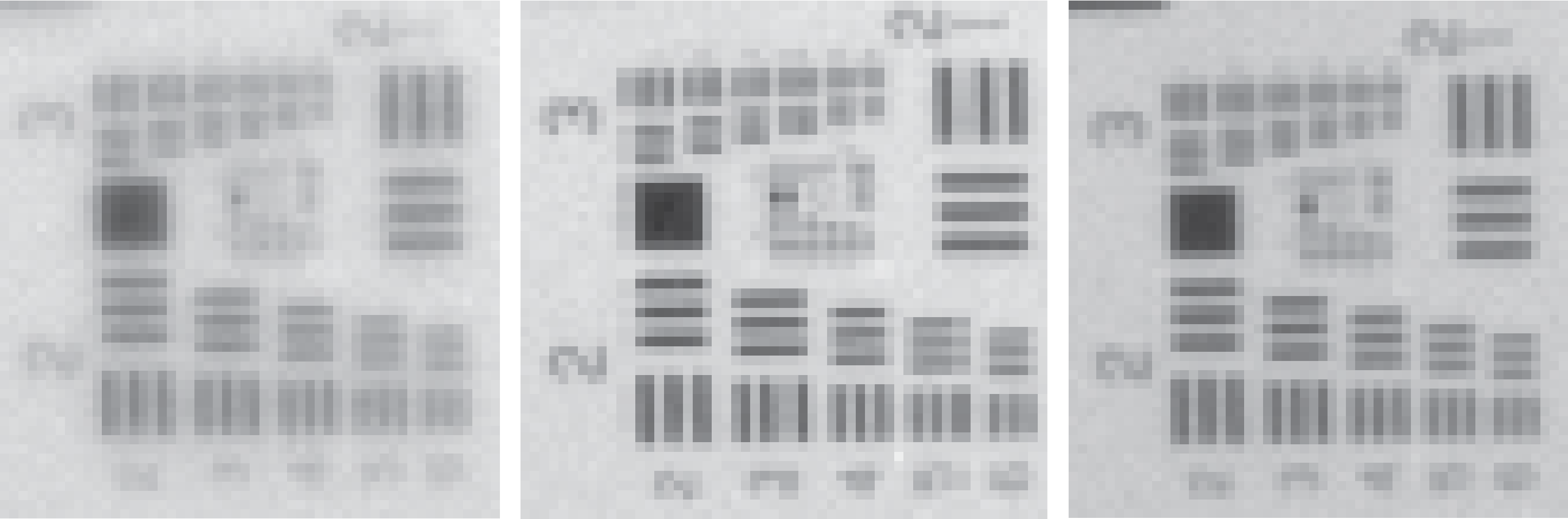}
    \caption{From left to right: UV, violet, and IR images of the USAF 1951 bar target, taken with the 12 mm lens. The full-sized images have been cropped so that only Groups 2 and 3 are visible. These images are focused with individual focus settings for each wavelength. Group 2, Element 3 is better resolved with these wavelength-specific focus settings than with common focus.}
    \label{fig:best focus}
\end{figure}

When testing focus, we used a USAF 1951 bar target to examine the approximate resolution of the system. Group 2, Element 3 of this bar target represents a spatial frequency of five line pairs per millimeter, which corresponds to our desired 0.1 mm/pixel resolution. Though some bands achieve better contrast than others, at best focus all bands resolve the lines in this group. With the common focus setting, UV, violet, and IR barely achieve this resolution. Fig. \ref{fig:common focus} shows the focus target imaged at all wavelength bands using amber focus. Fig. \ref{fig:best focus} shows the focus target imaged with best-focus settings for UV, violet, and IR.

The 50 mm lens is similarly wavelength dependent, though it requires different focus settings than the 12 mm lens. We found that focusing the 50 mm lens under green light creates sufficient focus for all bands except violet, deep red, and IR. Deep red requires its own focus setting for the 50 mm lens, while violet and IR can be focused with the same setting. 

\begin{figure}
    \centering
    \includegraphics[scale=0.8]{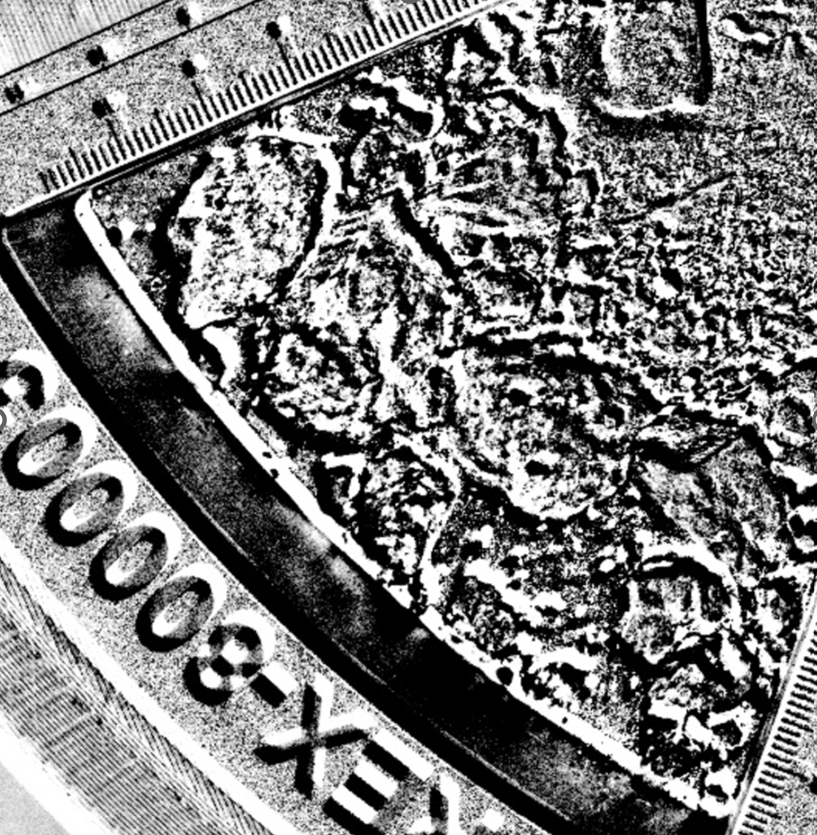}
    \includegraphics[scale=0.7]{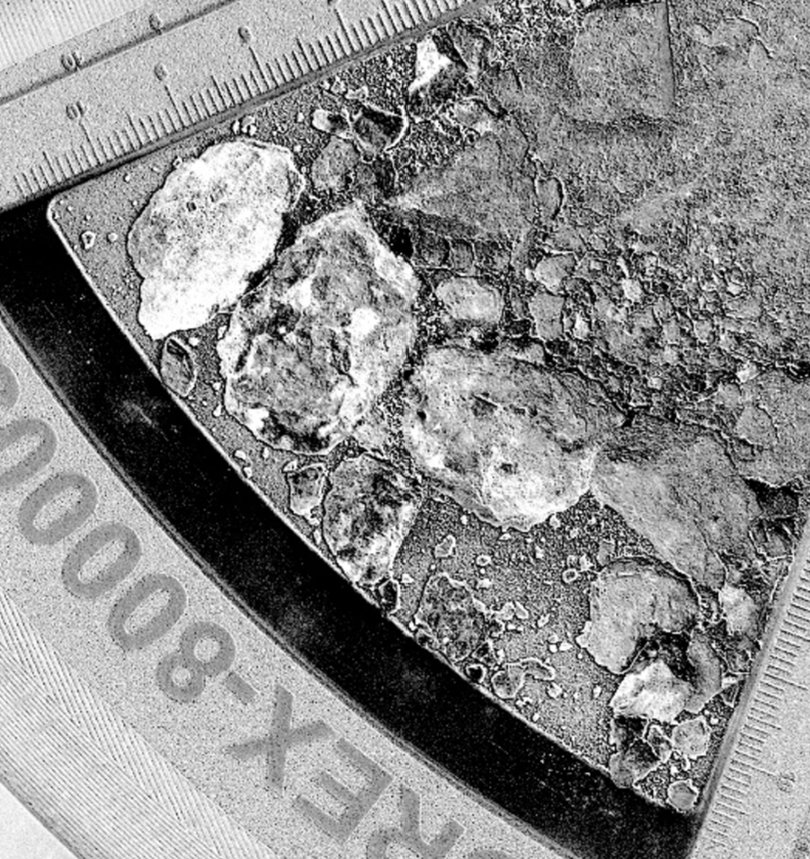}
    \caption{Left: IR/amber band ratio image of the same analog sample shown in Fig. \ref{fig:examples} and \ref{fig:zoom_example}, using an unregistered IR best-focus image. Artifacts caused by the difference in focal length between the IR and amber images are visible. Right: IR/amber band ratio image using an IR best-focus image registered to the amber image. Registration has eliminated the artifacts and resolved small features in the analog sample.}
    \label{fig:registration}
\end{figure}

Changing focus between images shifts the lens's focal length and by extension the images' pixel scale, which causes visible differences among images taken at different focus settings. To reduce the difference in pixel scale, we apply an algorithm that registers the UV, violet, and IR images in a sequence to the amber images for the 12 mm lens. The algorithm registers violet, deep red, and IR images to the green image for the 50 mm lens. Registration greatly reduces artifacts and improves the visibility of features in the field of view. Fig. \ref{fig:registration} compares IR/amber band ratios when using a non-registered vs. a registered IR image.  

\subsection{Pixel Scale and Distortion}
\begin{figure}
    \centering
    \includegraphics[scale = 0.05]{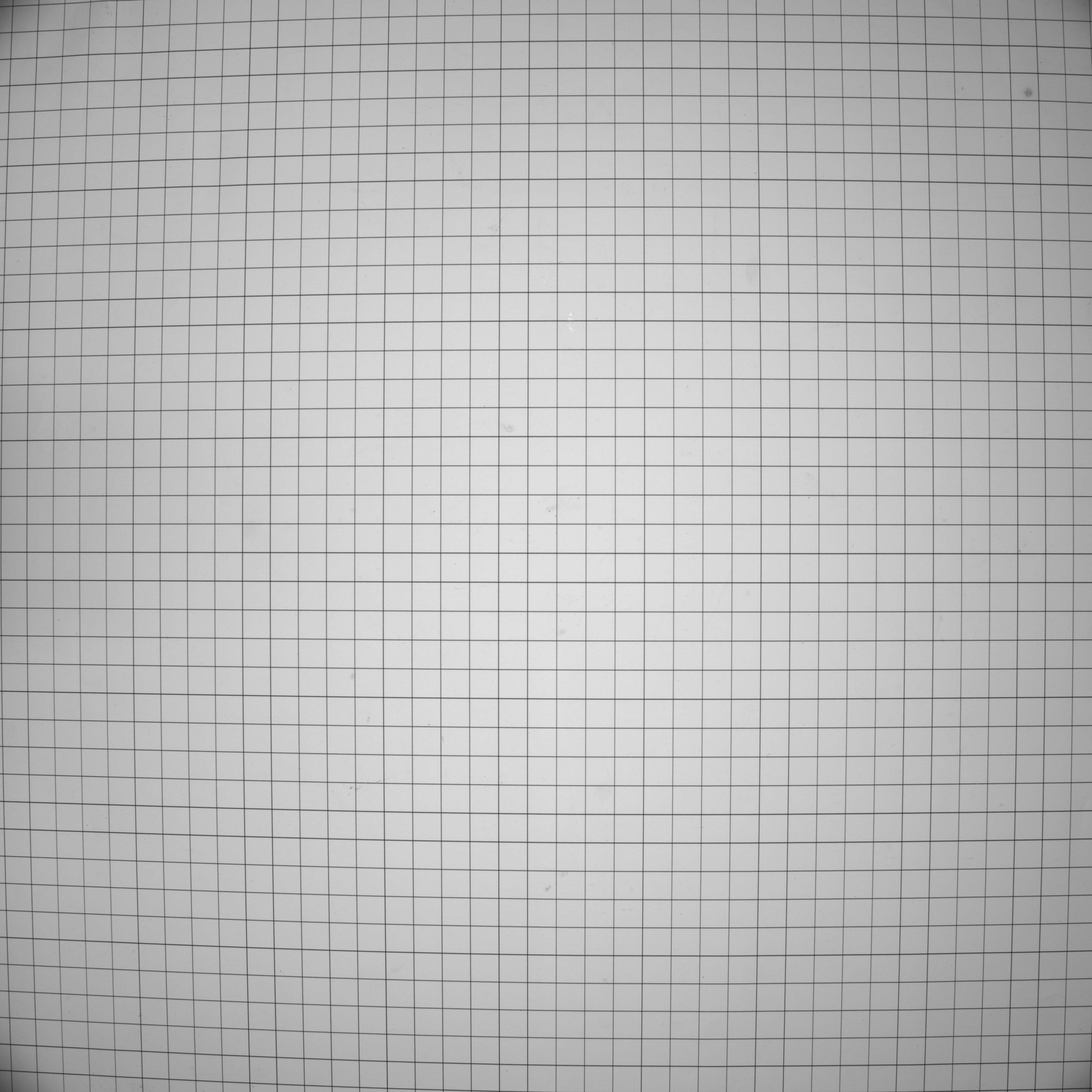}
    \caption{A grid of 1 cm squares imaged with the 12 mm lens. Distortion is visible toward the corners of the image.}
    \label{fig:dist12mm}
\end{figure}

\begin{figure}
    \centering
    \includegraphics[scale=0.5]{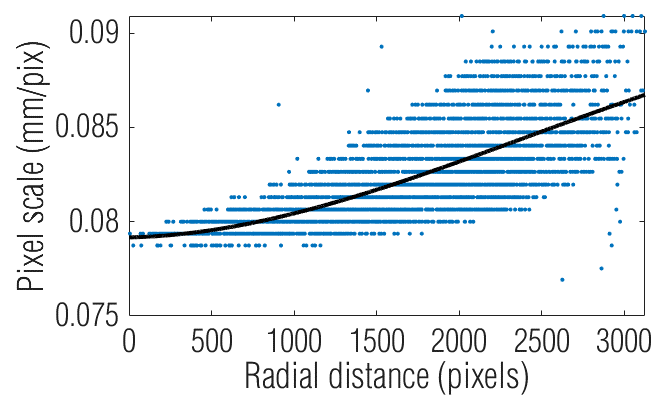}
    \caption{A third order polynomial fit to a scatter plot comparing pixel scale to radial distance from the image center.}
    \label{fig:distplot}
\end{figure}

QRIS images show angular and geometric distortion due to both optical distortion in the lens and the geometry of imaging a wide, 60\degree  field of view. Fig. \ref{fig:dist12mm} shows a QRIS image of a 1 cm grid taken with the 12 mm lens. Bowing of the grid is visible further from the center of the image. The grid images were coarsely analyzed to produce an estimate of the distortion by measuring the separation between grid lines. We assume that the distortion (both optical and geometric) is radially symmetric and we fit a third order polynomial to the data, shown in Fig. \ref{fig:distplot}. There is a $\approx$10\% increase in effective pixel size from center to corner of the field of view. We do not attempt to correct this distortion, as QRIS prioritizes reflectance and spectral data rather than spatial data. We document the distortion so that it can be taken into consideration when QRIS images are analyzed. 

\subsection{Tray Shadows and Reflections}
\begin{figure}
    \centering
    \includegraphics[scale=0.05]{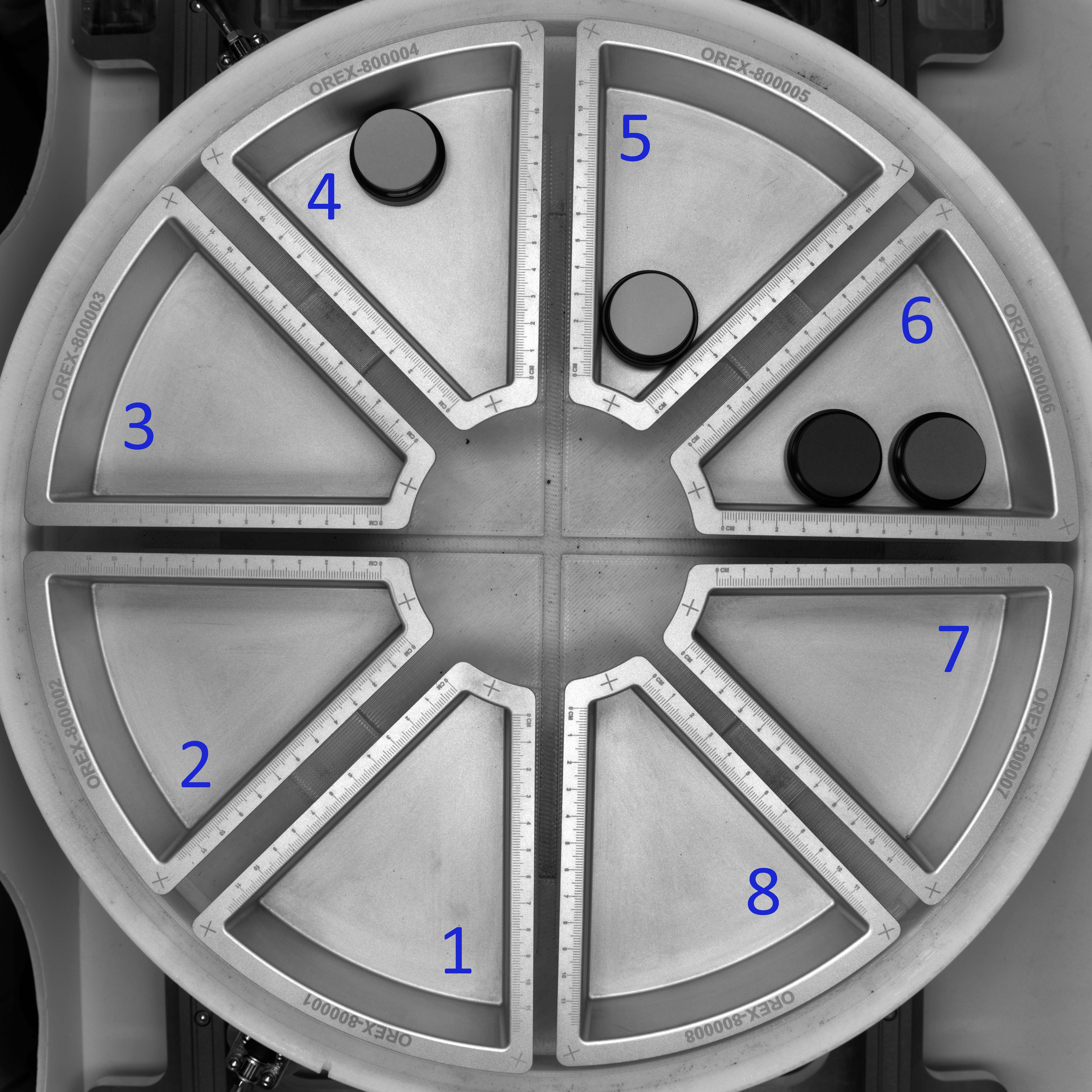}
    \caption{A broadband image of the empty aluminum sample trays, taken at JSC during a rehearsal in April 2023. The visible shadow along the lower wall of Tray 3 is plotted in Fig. \ref{fig:deeptrayplots}. Some Spectralon reflectance standards were placed in the trays to observe brightness variation on a surface with constant, known reflectance. Tray numbers are labeled.}
    \label{fig:emptytraysapril}
\end{figure}

\begin{figure}[h!]
    \centering
    \includegraphics[scale=0.35]{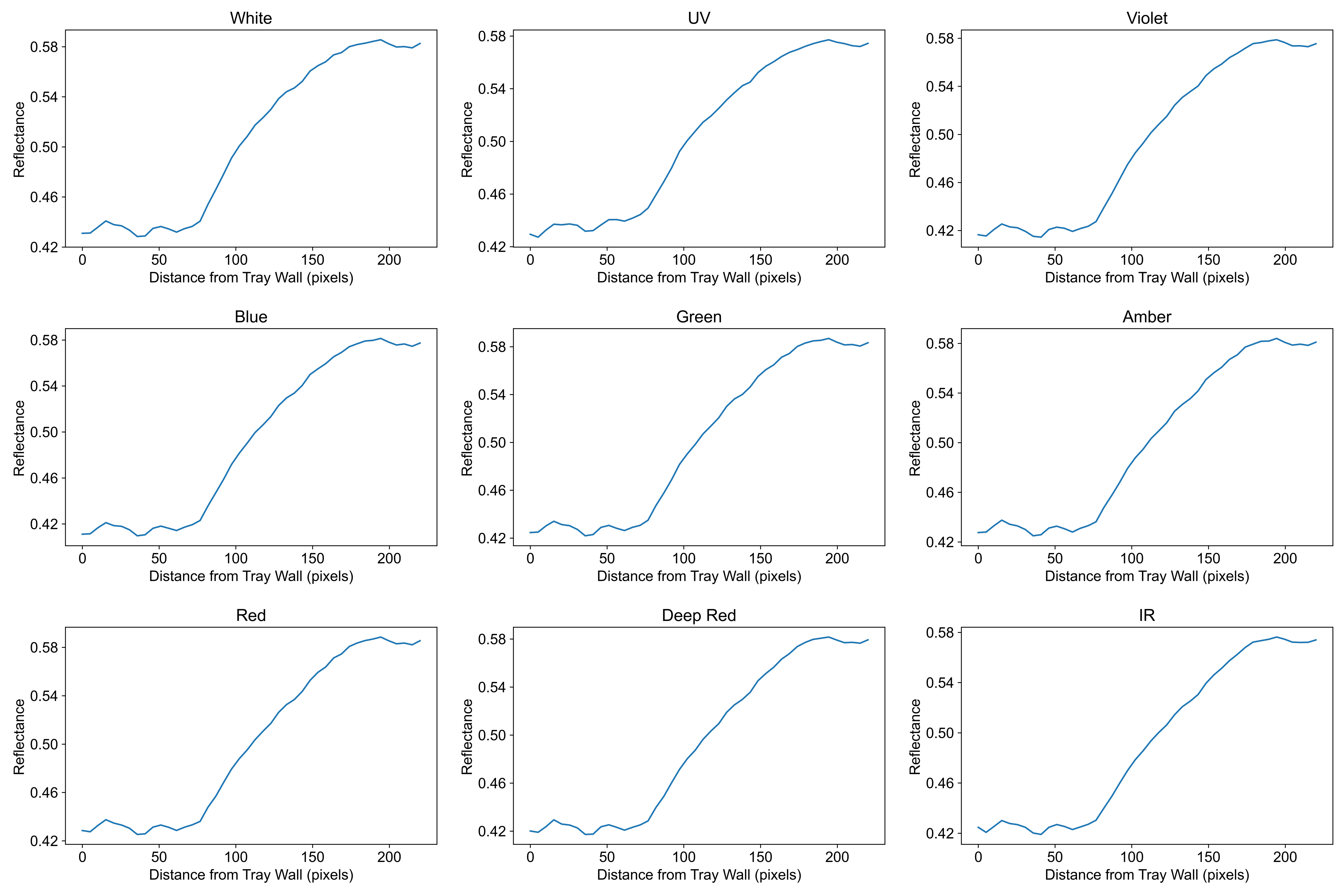}
    \caption{Plots of signal vs. distance in pixels from the Tray 3 wall, for each wavelength band. Signal values are calculated by averaging the signals in a series of thin lines (2 pixels wide) perpendicular to the tray wall. The fact that all wavelength bands show a similar signal pattern indicates that illumination diffusion (Section 2.3) is effective.}
    \label{fig:deeptrayplots}
\end{figure}

Fig. \ref{fig:emptytraysapril} shows the eight wedge-shaped sample trays, empty except for four Spectralon standards. Because they are designed to contain the sample rather than to facilitate imaging, the walls of the trays cast shadows and reflections into the trays, which could affect the perceived reflectance of regolith near the tray edges. Fig. \ref{fig:deeptrayplots} shows the change in signal as distance from the wall of Tray 3 increases. The lower wall of Tray 3 casts a visible shadow, and the signal is much lower within about 150 pixels of the tray's edge. Approximating the pixel scale as 0.1 mm/pixel, we can conclude that the tray is significantly shadowed within about 1.5 cm of the wall. However, this is a worst-case scenario caused by the position of this tray in relation to the lights. Not all trays are shadowed to this extent. 


The extent of tray shadows and reflections on the Bennu sample  depends on the specific topology of the regolith in each tray, which cannot be known until the sample is poured. Therefore, we do not attempt to correct the shadows or highlights as part of QRIS calibration. Nonetheless, when interpreting QRIS images, it is important to be aware that sample near the edges of the trays may appear brighter or darker than its true reflectance.

\subsection{Data Collection}
When imaging the Bennu sample, QRIS collects images with a range of exposure times (50-9999 ms, \S 3.4) for every wavelength band and lighting configuration. A full imaging sequence with the nominal 12 mm lens includes common focus images for all bands and best-focus images for UV, violet, and IR. A full imaging sequence is collected while the sample is in the TAGSAM head (e.g., Fig. 3, left), and another full sequence is taken after the sample is poured into the trays (e.g., Fig. 3, right). Once processed, the full sequence of tray images becomes the primary QRIS data product. Additional imaging sequences may be taken with the 50 mm lens if high-resolution data for regions of interest are needed.

\subsection{Data Products and Metadata}
Every nominal QRIS imaging sequence produces 27 Level-2 images, including one-sided illumination images. As previously stated, Level-2 images are saved as 32-bit TIFF files in units of reflectance. QRIS data products will be delivered to the Sample Analysis Micro-Information System (SAMIS) \citep{bennett2022}, where they can be accessed by the OSIRIS-REx team. The data uploaded to SAMIS will be archived in the publicly available AstroMat repository.

Following each imaging sequence, we will deliver two data products to SAMIS: (i) a raw data collection that includes all Level-0 images and (ii) a processed data collection that includes all Level-2 images along with Level-3 analysis products (discussed further below). Each data collection comprises two sets of images: one set for one-sided illumination, and one set for two-sided illumination. 

Data delivered to SAMIS will be accompanied by YAML-formatted metadata files of varying granularity. These files are included at the session level, containing metadata that apply to all of the images acquired in a single imaging sequence; at the data product level, containing metadata common to all the images in a single collection; and at the individual image level, containing QRIS settings used for each Level-0 and Level-2 image. The image-level metadata files are written automatically when images are collected and processed. Raw, Level-0 image metadata fields are listed in Table \ref{tab:metadata}. Processed, Level-2 image metadata fields are listed in Table \ref{tab:metadata2}.    

\begin{table}[h!]
\centering
\begin{tabular}{|p{4.5cm}|p{7cm}|}
     \hline \textbf{Metadata Field} &  \textbf{Description}\\ \hline
     Time Taken & Time of image collection. Formatted same as time in file names. \\ \hline
     Exposure Time & Exposure time in milliseconds \\ \hline
     Color & LED band name\\ \hline
     Configuration & Lighting configuration. A value of 1 means the right LED is illuminated, 2 means the left LED is illuminated, 3 means both sides are illuminated. \\ \hline
     Brightness & LED brightness setting. \\ \hline
     Focal Length & Focal length of lens used. \\ \hline
     Notes & Optional, additional information about imaging conditions. \\ \hline
     
     \end{tabular}
     
     \caption{Field descriptions for metadata files that accompany each raw QRIS image.}
     \label{tab:metadata}
\end{table}
\begin{table}[h!]
\centering
\begin{tabular}{|p{4.5cm}|p{7cm}|}
     \hline \textbf{Metadata Field} &  \textbf{Description}\\ \hline
     Color & LED band name. \\ \hline 
     Configuration & Lighting Configuration, see Table. \ref{tab:metadata}\\ \hline 
     Nonlinearity Parameters & Model parameters used in nonlinearity correction. \\ \hline 
     Flat file & File name of the flat image used for flat field correction in the calibration pipeline.  \\ \hline

     \end{tabular}
     
     \caption{Field descriptions for metadata files that accompany each Level-2 QRIS image.}
     \label{tab:metadata2}
\end{table}

\begin{figure}
    \centering
    \includegraphics[scale=0.05]{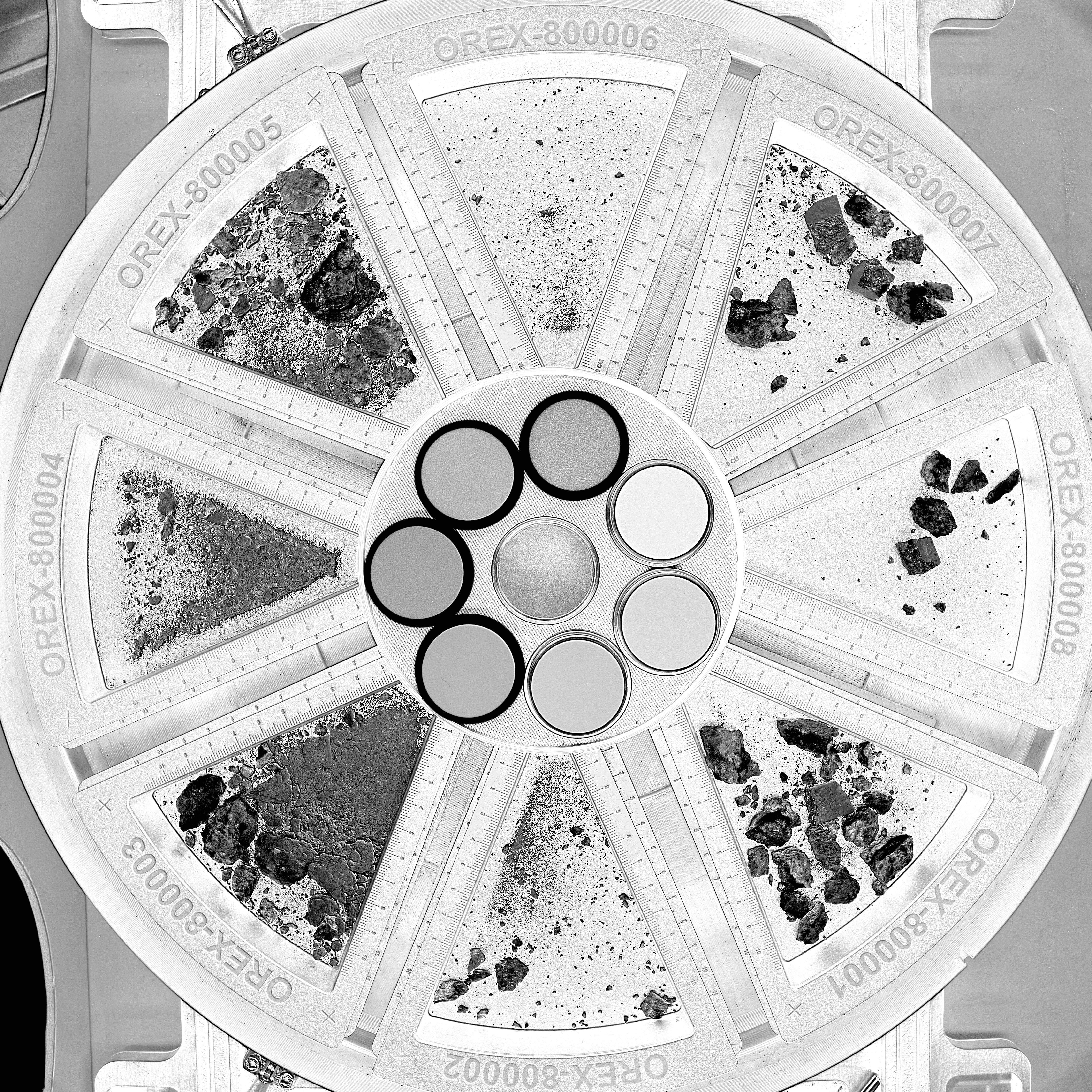}
    \includegraphics[scale=0.05]{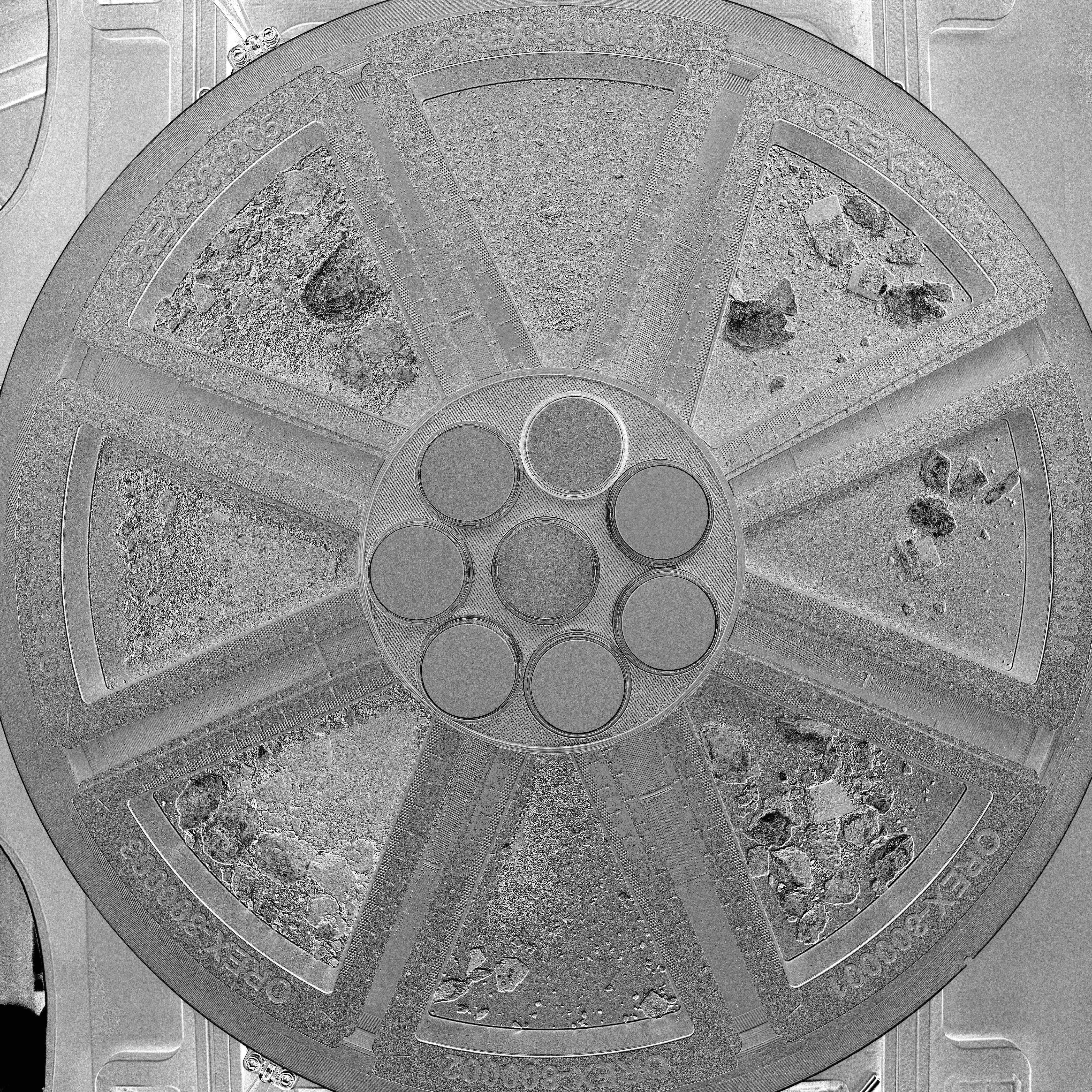}
    \caption{Left: Blue to infrared reflectance ratio for the analog sample used at the March 2023 rehearsal. Right: Green band depth map for the same analog sample. These images, and other similar products, were used to distinguish the lithologies in the analog sample.}
    \label{fig:boi_sopie}
\end{figure}

The Level-2 images can be used to create Level-3 analysis products such as band ratio and band depth maps, spectra of individual stones, and 3-channel color images. The following analysis products are delivered to the OSIRIS-REx science team as part of the processed data collection uploaded to SAMIS:  
\begin{center}
\begin{itemize}
  \item 3-channel RGB image
  \item 3-channel image with IR/green band ratio in the red channel, deep red band depth in the green channel, and blue/green band ratio in the blue channel (analogous to the false-color map in \cite{dellarefl})
  \item Amber band depth
  \item Deep red band depth
  \item Green band depth
  \item Blue/green band ratio
  \item Blue/IR band ratio
  \item IR/green band ratio
  \item IR/red band ratio 
  \item UV/green band ratio
  \item UV/IR band ratio
\end{itemize}
\end{center}
These products can elucidate features of the sample that are not visible in monochromatic reflectance images. Fig. \ref{fig:boi_sopie} shows two examples of such data products: a blue/IR band ratio map and a green band depth map used in characterizing the analog sample first pictured in Fig. \ref{fig:examples}. Fig. \ref{fig:boi_sopie} highlights differences between stones that are not obvious from the RGB images in Fig. \ref{fig:examples}. The features of the analog sample revealed by band ratios, band depths, and spectra were vital to the science team's success in distinguishing its seven lithologies.

\section{Results and Discussion}
In October 2023, QRIS imaged the recently delivered Bennu sample in the glovebox at JSC. Due to challenges opening the TAGSAM head, no imaging of the sample in the TAGSAM was performed. Moreover, QRIS images did not include the entire sample, only the particles that could be recovered from the TAGSAM by manually lifting its mylar flap, which were placed in four wedge-shaped trays. When images taken with the 12 mm lens were processed by our calibration pipeline, large regions of the calibrated reflectance images displayed negative reflectance values. This non-physical result revealed a flaw in our calibration method. We also observed artifacts created by light reflecting off the reflective glovebox walls, the glass glovebox window, and the QRIS gantry. The glovebox window is Schott Amiran glass, which is anti-reflection--coated to have $\approx$98\% transmittance (single pass) at visible wavelengths \citep{sch}. For QRIS, this manifests as 96\% transmittance (double pass) because the light passes through the glass twice. In this manuscript, we refer to transmittance by the double pass value. However, visual inspection and subsequent measurements indicate that the glovebox window transmittance is lower than advertised which may contribute to the calibration issues, as discussed in Section 5.4.

To the naked eye, most of the sample, though very dark, appeared brighter than the darkest (2\%) reflectance standard. However, QRIS DN values for the darkest standard were brighter than much of the sample, suggesting the presence of spatially variant stray light brightening the reflectance standards and thus lowering the calculated reflectance values of dark material. The more extreme wavelengths, such as violet and IR, tended to have the most negative values; small particles in violet images had reflectances down to -3\%. 

The Bennu sample imaging procedure included imaging with the 12 mm and 50 mm QRIS lenses. Due to the relatively small field of view of the 50 mm imaging sequences, the reflectance standards and the sample are imaged separately, but at approximately the same location within the camera’s field of view. Imaging with the 50 mm focal length lens could therefore lessen the effect of a spatially dependent artifact brightening the standards, thus improving the calibration. Unfortunately, the 50 mm calibration still showed considerable non-physical reflectance values. This implies that the stray light is not only spatially variant but also scene-dependent. That is, when changing the target from the calibration standards to the Bennu sample, they presented a different scene and therefore imposed a different form of stray light in the images.

\subsection{Calibration attempts}
Because the reflectance standards appeared, by inspection, to have higher than expected signal (implying additional illumination on them), we initially attempted to correct the issue by estimating and removing the additional signal. To estimate the additional signal, we fit a line to the signal-reflectance relationship for the 2\% and 5\% standards in HDR L1 images and calculated the y-intercept. The y-intercept is the theoretical signal value for a 0\% reflectance target and therefore represents the additional signal in that region of the image. To remove the additional signal, we subtracted the y-intercept from the 2\% standard DN values and then proceeded with the QRIS calibration described in Section 3. While this method did make most (all except very deep shadows) reflectance values positive, subsequent spectral analyses indicated that it did not produce accurate absolute reflectance values.

We attempted to make this approach more robust by adding a step to the calibration pipeline in which we fit a line to the signal-reflectance relationship for each band and exposure time. For each image, we subtracted the calculated y-intercepts and continued with calibration. In principle, this applies the correction described above to all images, rather than just those imaging very dark (2-5\%) materials. However, this method boosted sample reflectance to unreasonably high values. For example, the violet image showed dust reflectance values of around 6\% and large stone reflectance values of about 10\%, brighter than expected for all but the very brightest Bennu material. These results do not necessarily demonstrate that the reflectance standards were over-corrected; they may instead demonstrate that the stray light signal was removed from the standards but not the sample, thus increasing the calculated reflectance of the sample. This is yet another indication that the stray light is both spatially variant and scene-dependent. 

As mentioned above, artifacts from reflections off the interior of the glovebox were abundant in QRIS images. We were not able to rehearse QRIS imaging within the chamber, so could not anticipate how the artifacts would affect reflectance measurements. The glass glovebox window created a large reflection of the camera lens itself on the sample in the QRIS data. This lens reflection was brightest at long wavelengths, and the wavelength-dependent nature of this artifact meant it could not be eliminated by taking band ratios. During imaging, we noted multiple instances where reflections off the chamber, the QRIS gantry, or lights from elsewhere in the cleanroom created artifacts within the images. There were likely other reflections that were not as easily identifiable as artifacts. Due to these unknown and randomly distributed artifacts, it was impossible to confidently determine whether variation in brightness between stones was the result of artifacts or real variations in sample reflectance.

\subsection{Stray Light Sources}
Because these ad hoc methods to correct the additional signal were unsuccessful, we attempted to directly quantify the stray light instead. This would allow us to put bounds on the magnitude of the effects, even if we could not remove them. We did not have access to the JSC glovebox before or after QRIS imaging of the Bennu sample, so we were not able to explore stray light reflections within the glovebox. However, we did explore stray light sources using the QRIS prototype installed at the University of Arizona (UA). The prototype glovebox does not match the final glovebox (the prototype was manufactured first), so we cannot directly explore glovebox reflection. Nevertheless, the prototype uses an identical lens and detector, so we can investigate in-lens sources of stray light that are inherent to the camera. 

\begin{figure}
    \centering
    \includegraphics[scale=0.5]{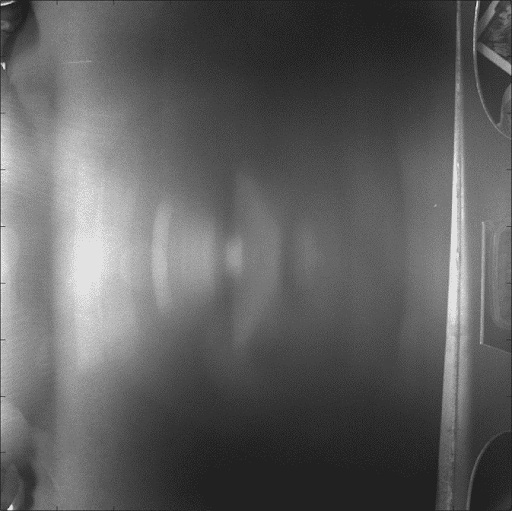}
    \caption{Stray light pattern produced by an out-of-field light source (a green LED strip). This image was taken with a 10 s exposure to emphasize these faint artifacts.}
    \label{fig:oofsl}
\end{figure}

Stray light can originate inside or outside of the camera field of view. To quantify the magnitude of out-of-field stray light, we placed a bright source outside the camera field of view. Photons from bright sources can reflect off internal lens surfaces (e.g., the inside of the lens barrel) and are imaged on to the detector by downstream optical surfaces. In laboratory testing, we have confirmed that out-of-field light sources can produce stray light artifacts, such as those shown in Fig. \ref{fig:oofsl}. However, these images showed that the stray light signal was 10,000 – 100,000 times darker than the signal that results from imaging the light source directly (in the field of view). Though this was not a rigorous stray light test (which would move the light source over a range of angles in two axes), manual movement of the light source suggested that the worst-case stray light was at least 10,000 times darker than the source. As such, we conclude that out-of-field stray light did not significantly affect QRIS data.

\begin{figure}[!h]
    \centering
    \includegraphics[scale=0.06]{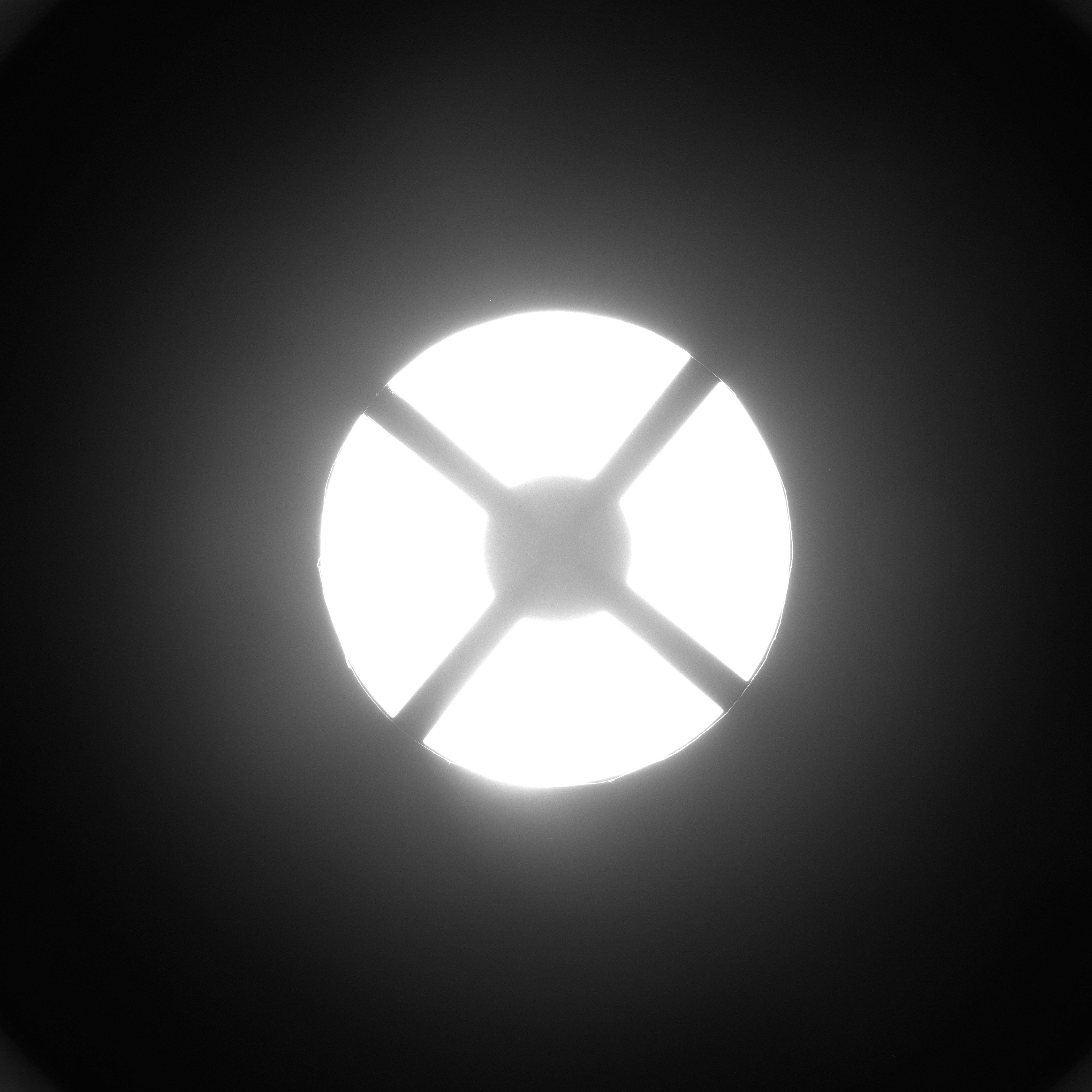}
    \caption{A contrast target illuminated from beneath by a blue LED strip. Holes in the mask are covered with the same white diffusing glass used in the QRIS light boxes. This image was taken with an exposure time of 500 ms. The region of the mask around the bright window is visibly brightened by stray light. This image was taken in the model glovebox at the University of Arizona.}
    \label{fig:mask_centered}
\end{figure}

To test the magnitude of in-field stray light, we imaged a contrast target: a transparent mask of white diffusing glass illuminated from behind by the QRIS LEDs. This is a different illumination geometry than typical QRIS imaging, which illuminates the target from above. This alternate geometry presents a scene with large contrast between the illuminated and masked regions. Figure \ref{fig:mask_centered} shows an example image taken of this experiment. 

One source of ‘stray light’ tested with this experiment is due to a system optical point spread function (PSF) that has broad wings. As a result, a bright source can have a glow around it. In a contrast target such as the one used in these tests, the wings of the PSF can extend into the dark center of the target and fill it with additional signal. In these test images, the signal in the central region of the target had $\sim$0.5-1\% of the signal from the contrast target.

\begin{figure}[!h]
    \centering
    \includegraphics[scale=0.06]{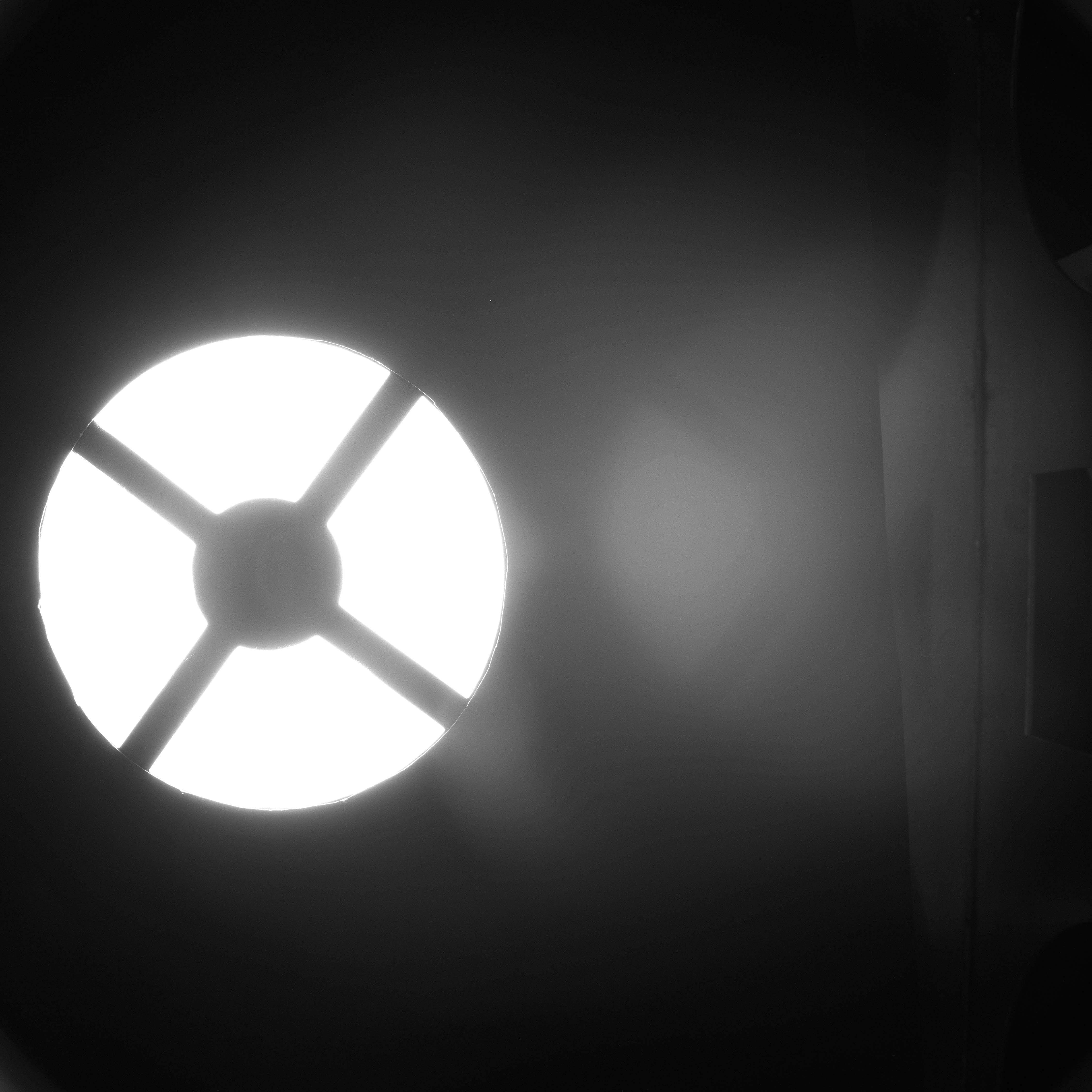}
    \caption{The same mask pictured in Figure \ref{fig:mask_centered}, with the target pushed toward the back of the model glovebox so that the bright region is off-axis, producing a ghost artifact to its right. The mask is illuminated from beneath by a blue LED strip and imaged with an exposure time of 500 ms.}
    \label{fig:mask_ghost}
\end{figure}

When we imaged the contrast target off-axis, we observed a ghost of the target, shown in Fig. \ref{fig:mask_ghost}. Ghosts are images of bright sources in the field of view that result from internal reflections (e.g. reflecting from the detector plane, to a lens surface, and back to the detector). By acquiring long exposures, we can measure the magnitude of the ghost relative to the source. In these test images, the ghost was as bright as 1\% of the contrast target.

In addition to ghosting, imaging a bright scene can produce stray light within the field of view from scattering off optical surfaces. In these data, in-field stray light manifests as a diffuse pattern in the middle of the field of view. The stray light pattern is circular, peaks in the middle of the field of view, and its magnitude depends on the brightness and location of targets in the scene. In these test images, the peak brightness of the in-field stray-light was approximately 1\% of the contrast target.

When the contrast target is imaged on-axis (Fig. \ref{fig:mask_centered}), these three effects are overlaid, along with any other forms of stray light that were not easily identified in off-axis images. The signal in the central region of the contrast target is $\approx$3\% of the signal from the target itself. 

We also imaged the contrast target through a pane of sample Schott Amiran glass with a luminous transmittance of $\approx$96\% to observe how a glass window would affect stray light. In-field stray light signal increased by a factor of up to 1.5 times when the glass was present. The glass window present during Bennu sample imaging may have worsened the effects of in-field stray light, thus brightening the dark standards and affecting the radiometric calibration.

Images of this contrast target are an extreme example, where a theoretically black target is adjacent to an extremely bright source. Imaging of the Bennu sample did not have as extreme a juxtaposition. However, the results suggest that a dark object (such as Bennu sample) surrounded by a bright scene (such as reflective metal surfaces) could have additional signal on the dark target that is as high as 1-3\% of the bright surroundings. Because the reflective metal (with reflectance values between 0.5 and 1) is $\approx$50$\times$ brighter than Bennu sample (with an expected reflectance of $\approx$2-4\%), the QRIS measurement of dark sample could be combined with stray light of similar magnitude.

\subsection{Insights from Rehearsal Data}
While we have established that the QRIS lenses have internal stray light, we have not been able to demonstrate that they are the source of the calibration failure during sample imaging. The final QRIS build was based on results from the third imaging rehearsal, conducted at JSC (Figs. \ref{fig:examples} and \ref{fig:zoom_example}). The data taken at this rehearsal did not have any indications of the calibration issues encountered during sample imaging. They do not contain reflection artifacts and contained very few negative or zero reflectance values. The analog sample used at this rehearsal was 3-4 times brighter than the Bennu sample, which could have masked the effect of stray light. In addition, the model glovebox used during this rehearsal did not have a glass window and was made of foam core (much less reflective than the polished metal interior of the cleanroom glovebox). The lack of a glass window during the rehearsal may have lessened the effect of stray light. As discussed in the previous section, the presence of glass with $\approx$96\% transmittance increased stray light by up to 1.5 times. As we will discuss in the following section, the glass used in the cleanroom glovebox likely has lower transmittance than expected, and therefore could intensify stray light more than the glass used in our tests at UA.

\subsection{Bennu sample imaging conditions}

\begin{figure}
    \centering
    \includegraphics[scale=0.7]{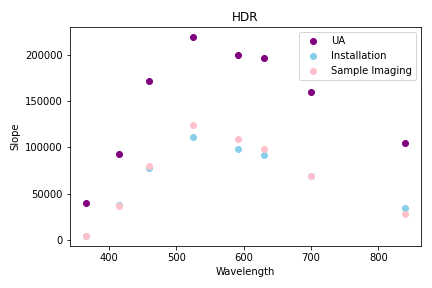}
    \caption{Slope of the signal-reflectance relationship at each band in HDR L1 images taken with no window (UA prototype), uncoated window (installation in cleanroom), and coated window (Bennu sample imaging in cleanroom).}
    \label{fig:slopes}
\end{figure}

\begin{figure}[!h]
    \centering
    \includegraphics[scale=0.35]{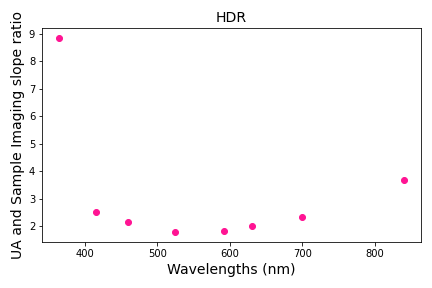}
    \includegraphics[scale=0.35
    ]{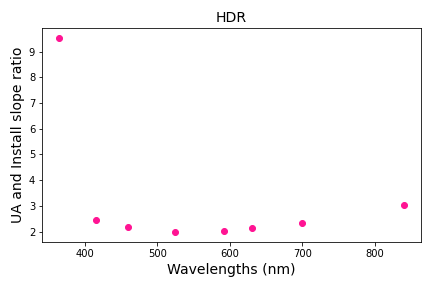}
    \includegraphics[scale=0.35]{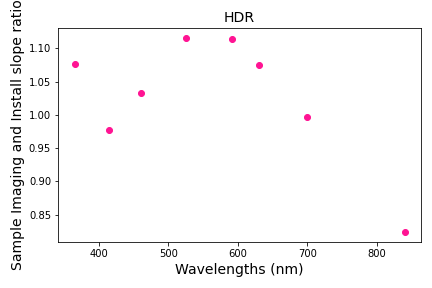}
    \caption{Ratios between signal-reflectance slopes for HDR images in each band gathered with no window (UA prototype), uncoated window (installation in cleanroom), and coated window (Bennu sample imaging in cleanroom). Left: Ratios between UA prototype and Bennu sample slopes. Center: Ratios between UA prototype and installation slopes. Right: Ratios between Bennu sample and installation slopes.}
    \label{fig:slope_ratios}
\end{figure}

QRIS will not have the opportunity to acquire diagnostic data at JSC to determine the impact of the glovebox window transmittance and stray light reflections within the glovebox on QRIS calibration. Instead, we compare data from Bennu sample imaging with data acquired during QRIS installation in the JSC cleanroom in July 2023. During QRIS installation, the JSC glovebox had not yet had its Amiran coated glass windows installed. Instead, the glovebox had uncoated glass windows. The transmittance of the glass was not measured, but we can estimate it using the average index of refraction of common uncoated optical glass ($n_{glass}\approx$1.5). We estimate the reflectivity of this average glass as 
\begin{align}
    R = 100*\frac{(n_{air}-n_{glass})^{2}}{(n_{air}+n_{glass})^{2}} = 4\%,
\end{align}
where R is reflectivity and the index of refraction for air is assumed to be 1. This reflection occurs at every glass/air boundary. Photons travelling from the light source to the target to the camera will experience four glass/air boundaries. Therefore, the total optical transmittance is $\approx$84\% (100\%-4*4\%). We further compare both JSC data sets to data acquired at UA with the QRIS prototype and no glovebox window. In all three datasets, we evaluate the relationship between the reflectance values of Spectralon standards and their measured DN signal values in QRIS images. This relationship is a measure of the illumination level of the scene. A steeper slope (higher signal to reflectance ratio) indicates a brighter scene. We compared the signal vs. reflectance relationship for data acquired during Bennu sample imaging at JSC, during QRIS installation at JSC, and in prototype testing at UA.  Fig. \ref{fig:slopes} shows the slopes at each band for HDR L1 images. Fig. \ref{fig:slope_ratios} visualizes the ratios between slopes in HDR L1 images.

We found that across the QRIS wavelength bands, images taken in the prototype glovebox had the highest signal-reflectance slopes, as expected because they have no loss of transmission through a window. The images taken of the Bennu sample at JSC through an Amiran glass window have the second highest DN-reflectance slopes. Images taken at QRIS installation through the uncoated window have the lowest slopes (for all bands except IR), which again matches expectation. However, if the Amiran glass used for Bennu sample imaging does indeed have a $\approx$96\% transmittance, it should produce slopes closer to the prototype (no window, 100\% transmittance) than to the uncoated glass window from installation ($\approx$ 84\% transmittance). Fig. \ref{fig:slopes} shows that slopes in images taken through the Amiran glass at sample imaging are much closer to uncoated installation glass slopes. This indicates that the Amiran glass used buring sample imaging has a lower transmittance (and therefore higher reflectivity) than expected. 

The slopes described above are calculated by fitting a line to the signal-reflectance relationship in the three data sets. These fits also produce y-intercept values. In principle, the y-intercepts should be very close to zero (i.e., zero reflectance should result in zero signal). In practice, this is not true. In these data the y-intercepts vary much more than expected and are representative of additional stray light in the scene. This is essentially the same analysis used in our initial attempts to correct the calibration (Section 5.1), where we removed the additional signal (estimated by the y-intercept). As noted in that section, that strategy did not improve the calibration. As such, this analysis cannot be used to create a correction, but it does confirm the behavior.
\begin{figure}
    \centering
    \includegraphics[scale=0.3]{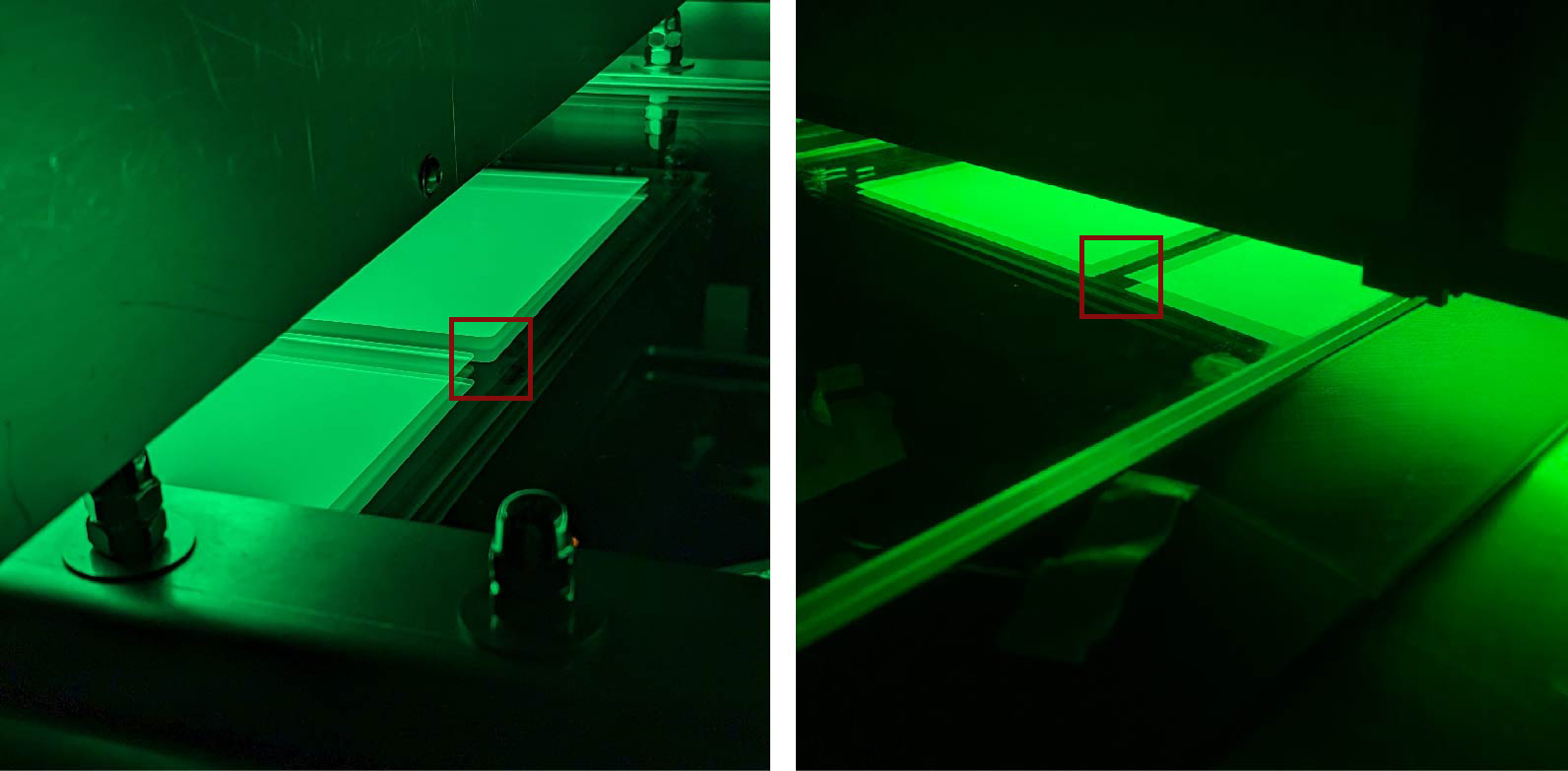}    
    \caption{Left: The reflections of green LEDs in the coated glass glovebox window at PE. Three reflections are seen, indicating that the window has an intermediate index of refraction boundary in the middle of the glass. Right: The reflection of green LEDs in the coated glass window added to the prototype glovebox at UA. Only two reflections are seen, which differs from PE and indicates a window with no intermediate boundary. The red squares emphasize the reflections in each image.}
    \label{fig:refl_layers}
\end{figure}
\begin{figure}
    \centering
    \includegraphics[scale=0.7]{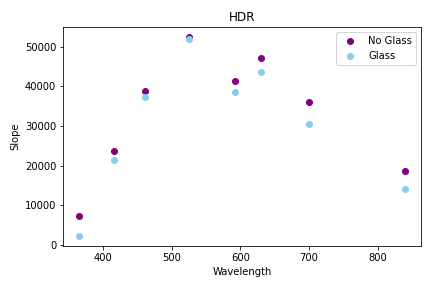}
    \caption{The signal-reflectance slope on prototype Spectralon reflectance standards imaged in the prototype glovebox at UA. At all bands, the slopes are slightly higher when the Amiran glass window, which has a transmittance of $\approx$96\%, is not present.}
    \label{fig:glass_comp}
\end{figure}

\begin{figure}
    \centering
    \includegraphics[scale=0.7]{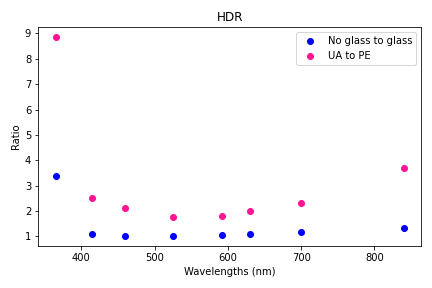}
    \caption{This plot shows in pink the ratio between the signal-reflectance slope on custom Spectralon standards imaged in the prototype glovebox at UA and in the cleanroom glovebox during sample imaging. The blue dots show the ratio between the slope on the prototype standards imaged at UA without an Amiran glass window and with an Amiran glass window.}
    \label{fig:ratio_comp}
\end{figure}

To further explore slope behavior, we imaged a prototype set of Spectralon reflectance standards in the UA prototype glovebox with and without a sample Amiran glass window. By comparing signal on bright standards with and without this window, we confirmed that this pane of sample Amiran glass has a transmittance of $\approx$96\%. Moreover, we noticed that the reflections of the LEDs off of the glass at UA were different than those during sample imaging at JSC (Fig. \ref{fig:refl_layers}); the latter shows three layers of reflection while the former glass shows only two. The difference between these reflections cannot be used to make quantitative conclusions about the transmittance of either glass, but it does demonstrate that the  coated glass in the cleanroom behaves differently from glass we know to be $\approx$96\% transmittance coated glass. We calculated the signal-reflectance slopes for the prototype standards in each band, with and without glass, plotted in Fig. \ref{fig:glass_comp}. Once again, we found that slopes were higher in images taken without the glass window. However, the slopes from these new images with glass at UA were much closer to the new UA slopes without glass than the sample imaging slopes were to the original UA slopes. Figure \ref{fig:ratio_comp} compares the ratio between slopes in images taken with and without glass at UA to the ratio between slopes in sample anylsis images and the original UA images.
\begin{figure}
    \centering
    \includegraphics[scale=0.7]{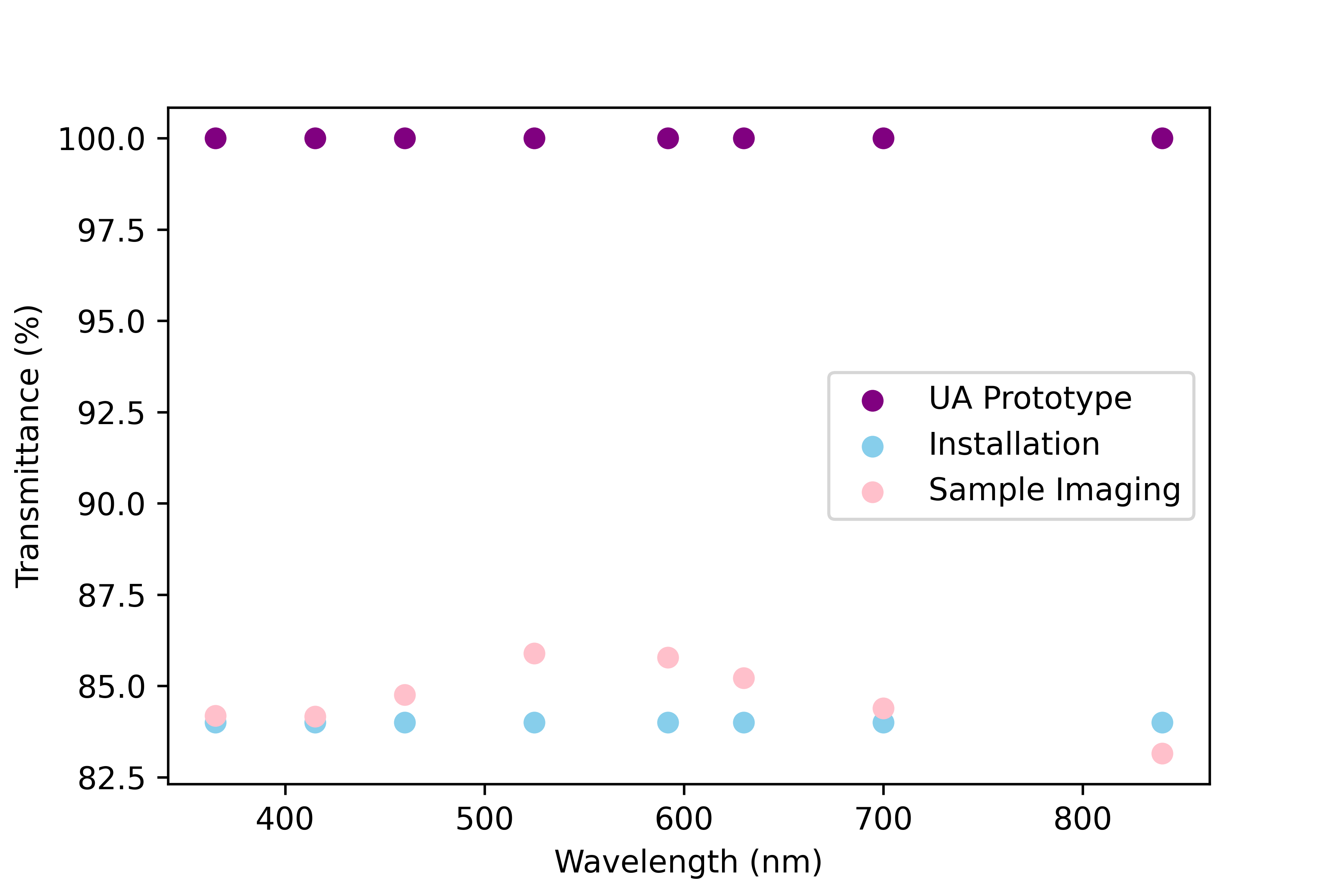}
    \caption{Calculated estimates of the cleanroom glovebox glass transmittance at each band assuming that installation transmittance is 84\% and that slope and transmittance are linearly related.}
    \label{fig:transmit}
\end{figure}

If we assume the relationship between slope and glass transmittance is linear, and that the scene and illumination conditions are identical, the ratios between slopes should be proportional to ratios between glass transmittances. The former assumption is likely true, the latter is not. The UA prototype data were acquired with the QRIS prototype, while installation and Bennu sample data were acquired with the final build. Though both have the same LED settings and number of LED strips, there could be slight differences between the two that would unfairly bias this comparison. For example, if the LED strips used during Bennu sample imaging were slightly dimmer than in the prototype, it would decrease the slope of the those data in a way that was not indicative of lower glass transmittance. Nonetheless, to first order, we believe the prototype and final builds are similar enough to justify a semi-quantitative comparison. If we define transmittance with no glass (UA prototype) as 100\% transmittance and assume 84\% transmittance for uncoated glass (as calculated above), we can linearly extrapolate between the two slope extremes to estimate the transmittance of the coated cleanroom glovebox glass. Fig. \ref{fig:transmit} plots the results, which estimates the transmittance of the coated cleanroom glovebox glass between 83 and 86\%, much lower than the advertised 96\% (two transmissions of 98\%). If true, we might expect that the 1.5$\times$ stray light amplification produced by 96\% transmittance glass would increase by an additional factor of 3-4$\times$ due to the more reflective coated cleanroom glovebox glass used at sample imaging. Given that the stray light inherent to the lens is high enough to compete with the signal from the Bennu sample, amplifying it by $>$3X would certainly corrupt the QRIS calibration. Even if we do not fully trust this estimation, the variation in signal-reflectance relationship for the prototype and cleanroom glovebox glass suggests that the latter has a lower transmittance than expected and would lead to increased stray light.

\section{Conclusion}
QRIS was designed to measure the reflectance of the Bennu sample in the glovebox in the OSIRIS\textendash REx cleanroom at JSC. We intended to use QRIS reflectance data to initially assess which of Bennu's major and minor lithologies are present in the sample. Because the two major lithologies observed by the OSIRIS-REx spacecraft at Bennu are very dark and only differ by a few percent reflectance, QRIS design requirements were driven by the need to measure small reflectance variations in dark material. 

QRIS collects images with a machine-vision camera and is illuminated by LED strips of  nine wavelength bands. The camera and lights are held in position about the glovebox window by a custom, adjustable gantry. The nominal QRIS lens has a focal length of 12 mm, which creates a field of view wide enough to image the entire sample. A 50 mm zoom lens is used for higher-resolution imaging of smaller regions. QRIS images are collected at multiple exposure times for each wavelength band. Because both QRIS lenses have wavelength-dependent focus, certain bands require additional imaging with separate focus settings. 

Raw QRIS data are processed by a Python calibration pipeline that applies detector nonlinearity correction, flat field correction, HDR stacking, and conversion to reflectance units. Accurate radiometric calibration of QRIS data is vital to our preliminary understanding of the nature of the sample. After processing, a complete QRIS imaging sequence produces a high-dynamic-range reflectance map of the sample for each band and lighting configuration. For bands that require individual focus settings, two images are produced: one taken with the band-specific focus setting, and one taken with a 'common' focus setting that matches that of other band images. The nominal dataset is a 'best focus' dataset that includes individually focused images for applicable bands and common focus images for all other bands, all registered to each other.

When viewing QRIS data, it is important to consider the effects of geometric distortion and the reflections and shadows cast on the sample by tray and glovebox walls. Though we do not attempt to correct these phenomena, their contribution to overall uncertainty must be taken into account.

Unfortunately, we encountered problems with QRIS imaging and calibration during Bennu sample imaging. Our raw data contained artifacts caused by reflections off the glovebox window and sides, and our calibration pipeline returned negative reflectance values. These issues were driven by the extremely dark sample, the unanticipated reflections in the glovebox, and stray light in our lens. The spectral data acquired by QRIS was not reliable enough to differentiate between distinct lithologies and identify material best suited for addressing mission science questions. Therefore, QRIS data were not used directly during sample characterization and allocation as originally planned. However, QRIS data were still useful for confirming conclusions drawn from other imaging and other analyses. Raw QRIS data, processed reflectance maps, and analytical products (such as band ratio and band depth maps) have been provided to the OSIRIS\textendash REx science team and will be made available to the general public via AstroMat. If publicly available QRIS data are analyzed in the future, it will be important to take into account the challenges we have outlined. The subpar performance of QRIS during Bennu sample imaging, in contrast to its successful performance at imaging rehearsals, highlights the importance of testing instruments in their intended environments whenever possible. If more rigorous QRIS testing had been performed on material as dark as Bennu in a highly-reflective environment, we may have been able to produce higher quality-QRIS data of the sample. 

 \section{Acknowledgments} 
This material is based upon work supported by NASA under Contract NNM10AA11C  and Award NNH09ZDA007O  issued through the New Frontiers Program. Bennu sample return is possible thanks to the efforts of the entire OSIRIS-REx team.   

QRIS images and derived products will be available via AstroMat (astromat.org). 

\section{Competing Interests}
The authors declare that they have no conflict of interest.

\bibliography{Fulford_qris_paper.bib}

\begin{thebibliography}{}

\bibitem[{A}llied Vision, 2023]{alliedvisionVimbaMachine}
{A}llied Vision (2023).
\newblock {V}imba {S}{D}{K} for machine vision cameras --- alliedvision.com.
\newblock \url{https://www.alliedvision.com/en/products/vimba-sdk/}.
\newblock [Accessed 03-08-2023].

\bibitem[Barton, 2019]{dmxenttecpro}
Barton, P. (2019).
\newblock Dmxenttecpro.
\newblock \url{https://pypi.org/project/DMXEnttecPro/}.
\newblock [Accessed 08-01-2023].

\bibitem[{Bennett} et~al., 2022]{bennett2022}
{Bennett}, C.~A., {Haenecour}, P., {Crombie}, M.~K., {Fitzgibbon}, M., {Ferro},
  A., {Hammond}, D., {McDonough}, E., {Westermann}, M.~M., {Barnes}, J.~J.,
  {Connolly}, H.~C., and {Lauretta}, D.~S. (2022).
\newblock {SAMIS: The OSIRIS-REx Sample Analysis Micro Information System}.
\newblock In {\em Goldschmidt 2022 Abstracts}, number 11647.

\bibitem[{Bierhaus} et~al., 2018]{tagsam}
{Bierhaus}, E.~B., {Clark}, B.~C., {Harris}, J.~W., {Payne}, K.~S., {Dubisher},
  R.~D., {Wurts}, D.~W., {Hund}, R.~A., {Kuhns}, R.~M., {Linn}, T.~M., {Wood},
  J.~L., {May}, A.~J., {Dworkin}, J.~P., {Beshore}, E., and {Lauretta}, D.~S.
  (2018).
\newblock {The OSIRIS-REx Spacecraft and the Touch-and-Go Sample Acquisition
  Mechanism (TAGSAM)}.
\newblock {\em Space Science Reviews}, 214(7):107.

\bibitem[Bradski, 2000]{opencv_library}
Bradski, G. (2000).
\newblock {The OpenCV Library}.
\newblock {\em Dr. Dobb's Journal of Software Tools}.

\bibitem[{DellaGiustina} et~al., 2020]{dellarefl}
{DellaGiustina}, D.~N., {Burke}, K.~N., {Walsh}, K.~J., {Smith}, P.~H.,
  {Golish}, D.~R., {Bierhaus}, E.~B., {Ballouz}, R.~L., {Becker}, T.~L.,
  {Campins}, H., {Tatsumi}, E., {Yumoto}, K., {Sugita}, S., {Deshapriya},
  J.~D.~P., {Cloutis}, E.~A., {Clark}, B.~E., {Hendrix}, A.~R., {Sen}, A., {Al
  Asad}, M.~M., {Daly}, M.~G., {Applin}, D.~M., {Avdellidou}, C., {Barucci},
  M.~A., {Becker}, K.~J., {Bennett}, C.~A., {Bottke}, W.~F., {Brodbeck}, J.~I.,
  {Connolly}, H.~C., {Delbo}, M., {de Leon}, J., {Drouet d'Aubigny}, C.~Y.,
  {Edmundson}, K.~L., {Fornasier}, S., {Hamilton}, V.~E., {Hasselmann}, P.~H.,
  {Hergenrother}, C.~W., {Howell}, E.~S., {Jawin}, E.~R., {Kaplan}, H.~H., {Le
  Corre}, L., {Lim}, L.~F., {Li}, J.~Y., {Michel}, P., {Molaro}, J.~L.,
  {Nolan}, M.~C., {Nolau}, J., {Pajola}, M., {Parkinson}, A., {Popescu}, M.,
  {Porter}, N.~A., {Rizk}, B., {Rizos}, J.~L., {Ryan}, A.~J., {Rozitis}, B.,
  {Shultz}, N.~K., {Simon}, A.~A., {Trang}, D., {Van Auken}, R.~B., {Wolner},
  C.~W.~V., and {Lauretta}, D.~S. (2020).
\newblock {Variations in color and reflectance on the surface of asteroid
  (101955) Bennu}.
\newblock {\em Science}, 370(6517):eabc3660.

\bibitem[{DellaGiustina} et~al., 2021]{DellaExo}
{DellaGiustina}, D.~N., {Kaplan}, H.~H., {Simon}, A.~A., {Bottke}, W.~F.,
  {Avdellidou}, C., {Delbo}, M., {Ballouz}, R.~L., {Golish}, D.~R., {Walsh},
  K.~J., {Popescu}, M., {Campins}, H., {Barucci}, M.~A., {Poggiali}, G.,
  {Daly}, R.~T., {Le Corre}, L., {Hamilton}, V.~E., {Porter}, N., {Jawin},
  E.~R., {McCoy}, T.~J., {Connolly}, H.~C., {Garcia}, J.~L.~R., {Tatsumi}, E.,
  {de Leon}, J., {Licandro}, J., {Fornasier}, S., {Daly}, M.~G., {Al Asad},
  M.~M., {Philpott}, L., {Seabrook}, J., {Barnouin}, O.~S., {Clark}, B.~E.,
  {Nolan}, M.~C., {Howell}, E.~S., {Binzel}, R.~P., {Rizk}, B., {Reuter},
  D.~C., and {Lauretta}, D.~S. (2021).
\newblock {Exogenic basalt on asteroid (101955) Bennu}.
\newblock {\em Nature Astronomy}, 5:31--38.

\bibitem[{Golish} et~al., 2020]{dathoncal}
{Golish}, D.~R., {Drouet d'Aubigny}, C., {Rizk}, B., {DellaGiustina}, D.~N.,
  {Smith}, P.~H., {Becker}, K., {Shultz}, N., {Stone}, T., {Barker}, M.~K.,
  {Mazarico}, E., {Tatsumi}, E., {Gaskell}, R.~W., {Harrison}, L., {Merrill},
  C., {Fellows}, C., {Williams}, B., {O'Dougherty}, S., {Whiteley}, M.,
  {Hancock}, J., {Clark}, B.~E., {Hergenrother}, C.~W., and {Lauretta}, D.~S.
  (2020).
\newblock {Ground and In-Flight Calibration of the OSIRIS-REx Camera Suite}.
\newblock {\em Space Science Reviews}, 216(1):12.

\bibitem[{Golish} et~al., 2021]{golish_map}
{Golish}, D.~R., {Shultz}, N.~K., {Becker}, T.~L., {Becker}, K.~J.,
  {Edmundson}, K.~L., {DellaGiustina}, D.~N., {Drouet d'Aubigny}, C.,
  {Bennett}, C.~A., {Rizk}, B., {Barnouin}, O.~S., {Daly}, M.~G., {Seabrook},
  J.~A., {Philpott}, L., {Al Asad}, M.~M., {Johnson}, C.~L., {Li}, J.~Y.,
  {Ballouz}, R.~L., {Jawin}, E.~R., and {Lauretta}, D.~S. (2021).
\newblock {A high-resolution normal albedo map of asteroid (101955) Bennu}.
\newblock {\em Icarus}, 355:114133.

\bibitem[{Hamilton} et~al., 2019]{Vicky2019}
{Hamilton}, V.~E., {Simon}, A.~A., {Christensen}, P.~R., {Reuter}, D.~C.,
  {Clark}, B.~E., {Barucci}, M.~A., {Bowles}, N.~E., {Boynton}, W.~V.,
  {Brucato}, J.~R., {Cloutis}, E.~A., {Connolly}, H.~C., {Donaldson Hanna},
  K.~L., {Emery}, J.~P., {Enos}, H.~L., {Fornasier}, S., {Haberle}, C.~W.,
  {Hanna}, R.~D., {Howell}, E.~S., {Kaplan}, H.~H., {Keller}, L.~P., {Lantz},
  C., {Li}, J.~Y., {Lim}, L.~F., {McCoy}, T.~J., {Merlin}, F., {Nolan}, M.~C.,
  {Praet}, A., {Rozitis}, B., {Sandford}, S.~A., {Schrader}, D.~L., {Thomas},
  C.~A., {Zou}, X.~D., {Lauretta}, D.~S., and {Osiris-Rex Team} (2019).
\newblock {Evidence for widespread hydrated minerals on asteroid (101955)
  Bennu}.
\newblock {\em Nature Astronomy}, 3:332--340.

\bibitem[Harris et~al., 2020]{numpy}
Harris, C.~R., Millman, K.~J., van~der Walt, S.~J., Gommers, R., Virtanen, P.,
  Cournapeau, D., Wieser, E., Taylor, J., Berg, S., Smith, N.~J., Kern, R.,
  Picus, M., Hoyer, S., van Kerkwijk, M.~H., Brett, M., Haldane, A., del
  R{\'{i}}o, J.~F., Wiebe, M., Peterson, P., G{\'{e}}rard-Marchant, P.,
  Sheppard, K., Reddy, T., Weckesser, W., Abbasi, H., Gohlke, C., and Oliphant,
  T.~E. (2020).
\newblock Array programming with {NumPy}.
\newblock {\em Nature}, 585(7825):357--362.

\bibitem[Janesick et~al., 1987]{janesick}
Janesick, J.~R., Elliott, T., Collins, S., Blouke, M.~M., and Freeman, J.
  (1987).
\newblock Scientific charge-coupled devices.
\newblock {\em Optical Engineering}, 26(8):692--714.

\bibitem[{Kaplan} et~al., 2020]{Kaplan2020}
{Kaplan}, H.~H., {Lauretta}, D.~S., {Simon}, A.~A., {Hamilton}, V.~E.,
  {DellaGiustina}, D.~N., {Golish}, D.~R., {Reuter}, D.~C., {Bennett}, C.~A.,
  {Burke}, K.~N., {Campins}, H., {Connolly}, H.~C., {Dworkin}, J.~P., {Emery},
  J.~P., {Glavin}, D.~P., {Glotch}, T.~D., {Hanna}, R., {Ishimaru}, K.,
  {Jawin}, E.~R., {McCoy}, T.~J., {Porter}, N., {Sandford}, S.~A., {Ferrone},
  S., {Clark}, B.~E., {Li}, J.~Y., {Zou}, X.~D., {Daly}, M.~G., {Barnouin},
  O.~S., {Seabrook}, J.~A., and {Enos}, H.~L. (2020).
\newblock {Bright carbonate veins on asteroid (101955) Bennu: Implications for
  aqueous alteration history}.
\newblock {\em Science}, 370(6517):eabc3557.

\bibitem[{Kaplan, H. H.} et~al., 2021]{kaplan2021}
{Kaplan, H. H.}, {Simon, A. A.}, {Hamilton, V. E.}, {Thompson, M. S.},
  {Sandford, S. A.}, {Barucci, M. A.}, {Cloutis, E. A.}, {Brucato, J.},
  {Reuter, D. C.}, {Glavin, D. P.}, {Clark, B. E.}, {Dworkin, J. P.}, {Campins,
  H.}, {Emery, J. P.}, {Fornasier, S.}, {Zou, X. D.}, and {Lauretta, D. S.}
  (2021).
\newblock Composition of organics on asteroid (101955) bennu.
\newblock {\em A\&A}, 653:L1.

\bibitem[{Kroger}, 2022]{VimbaPython}
{Kroger}, N. (2022).
\newblock {G}it{H}ub - alliedvision/{V}imba{P}ython: {A}llied {V}ision {V}imba
  {P}ython {A}{P}{I} --- github.com.
\newblock \url{https://github.com/alliedvision/VimbaPython}.
\newblock [Accessed 08-01-2023].

\bibitem[{Lauretta} et~al., 2022]{Dante2022}
{Lauretta}, D.~S., {Adam}, C.~D., {Allen}, A.~J., {Ballouz}, R.~L., {Barnouin},
  O.~S., {Becker}, K.~J., {Becker}, T., {Bennett}, C.~A., {Bierhaus}, E.~B.,
  {Bos}, B.~J., {Burns}, R.~D., {Campins}, H., {Cho}, Y., {Christensen}, P.~R.,
  {Church}, E.~C.~A., {Clark}, B.~E., {Connolly}, H.~C., {Daly}, M.~G.,
  {DellaGiustina}, D.~N., {Drouet d{\textquoteright}Aubigny}, C.~Y., {Emery},
  J.~P., {Enos}, H.~L., {Kasper}, S.~F., {Garvin}, J.~B., {Getzandanner}, K.,
  {Golish}, D.~R., {Hamilton}, V.~E., {Hergenrother}, C.~W., {Kaplan}, H.~H.,
  {Keller}, L.~P., {Lessac-Chenen}, E.~J., {Liounis}, A.~J., {Ma}, H.,
  {McCarthy}, L.~K., {Miller}, B.~D., {Moreau}, M.~C., {Morota}, T., {Nelson},
  D.~S., {Nolau}, J.~O., {Olds}, R., {Pajola}, M., {Pelgrift}, J.~Y., {Polit},
  A.~T., {Ravine}, M.~A., {Reuter}, D.~C., {Rizk}, B., {Rozitis}, B., {Ryan},
  A.~J., {Sahr}, E.~M., {Sakatani}, N., {Seabrook}, J.~A., {Selznick}, S.~H.,
  {Skeen}, M.~A., {Simon}, A.~A., {Sugita}, S., {Walsh}, K.~J., {Westermann},
  M.~M., {Wolner}, C.~W.~V., and {Yumoto}, K. (2022).
\newblock {Spacecraft sample collection and subsurface excavation of asteroid
  (101955) Bennu}.
\newblock {\em Science}, 377(6603):285--291.

\bibitem[{Lauretta} et~al., 2019]{dante2019}
{Lauretta}, D.~S., {Dellagiustina}, D.~N., {Bennett}, C.~A., {Golish}, D.~R.,
  {Becker}, K.~J., {Balram-Knutson}, S.~S., {Barnouin}, O.~S., {Becker}, T.~L.,
  {Bottke}, W.~F., {Boynton}, W.~V., {Campins}, H., {Clark}, B.~E., {Connolly},
  H.~C., {Drouet D'Aubigny}, C.~Y., {Dworkin}, J.~P., {Emery}, J.~P., {Enos},
  H.~L., {Hamilton}, V.~E., {Hergenrother}, C.~W., {Howell}, E.~S., {Izawa},
  M.~R.~M., {Kaplan}, H.~H., {Nolan}, M.~C., {Rizk}, B., {Roper}, H.~L.,
  {Scheeres}, D.~J., {Smith}, P.~H., {Walsh}, K.~J., {Wolner}, C.~W.~V., and
  {OSIRIS-REx Team} (2019).
\newblock {The unexpected surface of asteroid (101955) Bennu}.
\newblock {\em Nature}, 568(7750):55--60.

\bibitem[{Lauretta} et~al., 2021]{DanteBook}
{Lauretta}, D.~S., {Enos}, H.~L., {Polit}, A.~T., {Roper}, H.~L., and {Wolner},
  C. W.~V. (2021).
\newblock {Chapter 8 - OSIRIS-REx at Bennu: Overcoming challenges to collect a
  sample of the early Solar System}.
\newblock In {Longobardo}, A., editor, {\em Sample Return Missions: The Last
  Frontier of Solar System Exploration}, pages 163--194. Elsevier.

\bibitem[{Lauretta} et~al., 2023]{SAP}
{Lauretta}, D.~S., {Polit}, A.~T., {Connolly}, H.~C., {Grossman}, J., and
  {OSIRIS-REx Team} (2023).
\newblock {OSIRIS-REx Sample Analysis Plan - Revision 3.0}.
\newblock {\em arXiv [astro-ph.EP] 2308.11794.}

\bibitem[{Righter} et~al., 2023]{Righter2023}
{Righter}, K., {Lunning}, N.~G., {Nakamura-Messenger}, K., {Snead}, C.~J.,
  {MCQuillan}, J., {Calaway}, M., {Allums}, K., {Rodriguez}, M., {Funk}, R.~C.,
  {Harrington}, R.~S., {Connelly}, W., {Cowden}, T., {Dworkin}, J.~P.,
  {Lorentson}, C.~C., {Sandford}, S.~A., {Bierhaus}, E.~B., {Freund}, S.,
  {Connolly}, H.~C., and {Lauretta}, D.~S. (2023).
\newblock {Curation planning and facilities for asteroid Bennu samples returned
  by the OSIRIS-REx mission}.
\newblock {\em Meteoritics \& Planetary Science}, 58(4):572--590.

\bibitem[{Rizk} et~al., 2018]{rizk}
{Rizk}, B., {Drouet d'Aubigny}, C., {Golish}, D., {Fellows}, C., {Merrill}, C.,
  {Smith}, P., {Walker}, M.~S., {Hendershot}, J.~E., {Hancock}, J., {Bailey},
  S.~H., {DellaGiustina}, D.~N., {Lauretta}, D.~S., {Tanner}, R., {Williams},
  M., {Harshman}, K., {Fitzgibbon}, M., {Verts}, W., {Chen}, J., {Connors}, T.,
  {Hamara}, D., {Dowd}, A., {Lowman}, A., {Dubin}, M., {Burt}, R., {Whiteley},
  M., {Watson}, M., {McMahon}, T., {Ward}, M., {Booher}, D., {Read}, M.,
  {Williams}, B., {Hunten}, M., {Little}, E., {Saltzman}, T., {Alfred}, D.,
  {O'Dougherty}, S., {Walthall}, M., {Kenagy}, K., {Peterson}, S., {Crowther},
  B., {Perry}, M.~L., {See}, C., {Selznick}, S., {Sauve}, C., {Beiser}, M.,
  {Black}, W., {Pfisterer}, R.~N., {Lancaster}, A., {Oliver}, S., {Oquest}, C.,
  {Crowley}, D., {Morgan}, C., {Castle}, C., {Dominguez}, R., and {Sullivan},
  M. (2018).
\newblock {OCAMS: The OSIRIS-REx Camera Suite}.
\newblock {\em Space Science Reviews}, 214(1):26.

\bibitem[{Rozitis} et~al., 2020]{rozitis_2020}
{Rozitis}, B., {Ryan}, A.~J., {Emery}, J.~P., {Christensen}, P.~R., {Hamilton},
  V.~E., {Simon}, A.~A., {Reuter}, D.~C., {Al Asad}, M., {Ballouz}, R.~L.,
  {Bandfield}, J.~L., {Barnouin}, O.~S., {Bennett}, C.~A., {Bernacki}, M.,
  {Burke}, K.~N., {Cambioni}, S., {Clark}, B.~E., {Daly}, M.~G., {Delbo}, M.,
  {DellaGiustina}, D.~N., {Elder}, C.~M., {Hanna}, R.~D., {Haberle}, C.~W.,
  {Howell}, E.~S., {Golish}, D.~R., {Jawin}, E.~R., {Kaplan}, H.~H., {Lim},
  L.~F., {Molaro}, J.~L., {Munoz}, D.~P., {Nolan}, M.~C., {Rizk}, B.,
  {Siegler}, M.~A., {Susorney}, H.~C.~M., {Walsh}, K.~J., and {Lauretta}, D.~S.
  (2020).
\newblock {Asteroid (101955) Bennu's weak boulders and thermally anomalous
  equator}.
\newblock {\em Science Advances}, 6(41):eabc3699.

\bibitem[Rozitis et~al., 2022]{rozitis2022}
Rozitis, B., Ryan, A.~J., Emery, J.~P., Nolan, M.~C., Green, S.~F.,
  Christensen, P.~R., Hamilton, V.~E., Daly, M.~G., Barnouin, O.~S., and
  Lauretta, D.~S. (2022).
\newblock High-resolution thermophysical analysis of the osiris-rex sample site
  and three other regions of interest on bennu.
\newblock {\em Journal of Geophysical Research: Planets}, 127(6):e2021JE007153.

\bibitem[{Schott}, 2023]{sch}
{Schott} (2023).
\newblock {SCHOTT AMIRAN - Anti-Reflective Glass Technical Data Sheet}.
\newblock {\em schott.com}.

\bibitem[{Simon} et~al., 2020]{2020simon}
{Simon}, A.~A., {Kaplan}, H.~H., {Hamilton}, V.~E., {Lauretta}, D.~S.,
  {Campins}, H., {Emery}, J.~P., {Barucci}, M.~A., {DellaGiustina}, D.~N.,
  {Reuter}, D.~C., {Sandford}, S.~A., {Golish}, D.~R., {Lim}, L.~F., {Ryan},
  A., {Rozitis}, B., and {Bennett}, C.~A. (2020).
\newblock {Widespread carbon-bearing materials on near-Earth asteroid (101955)
  Bennu}.
\newblock {\em Science}, 370(6517):eabc3522.

\bibitem[Spring and Davidson, 2006]{spring_davidson_2006}
Spring, K. and Davidson, M. (2006).
\newblock Concepts in digital imaging technology: Dynamic range.
\newblock {\em Hamamatsu Learning Center: Dynamic Range}.

\bibitem[Virtanen et~al., 2020]{SciPy}
Virtanen, P., Gommers, R., Oliphant, T.~E., Haberland, M., Reddy, T.,
  Cournapeau, D., Burovski, E., Peterson, P., Weckesser, W., Bright, J., {van
  der Walt}, S.~J., Brett, M., Wilson, J., Millman, K.~J., Mayorov, N., Nelson,
  A. R.~J., Jones, E., Kern, R., Larson, E., Carey, C.~J., Polat, {\.I}., Feng,
  Y., Moore, E.~W., {VanderPlas}, J., Laxalde, D., Perktold, J., Cimrman, R.,
  Henriksen, I., Quintero, E.~A., Harris, C.~R., Archibald, A.~M., Ribeiro,
  A.~H., Pedregosa, F., {van Mulbregt}, P., and {SciPy 1.0 Contributors}
  (2020).
\newblock {{SciPy} 1.0: Fundamental Algorithms for Scientific Computing in
  Python}.
\newblock {\em Nature Methods}, 17:261--272.

\bibitem[{W}es {M}c{K}inney, 2010]{pandas}
{W}es {M}c{K}inney (2010).
\newblock {D}ata {S}tructures for {S}tatistical {C}omputing in {P}ython.
\newblock In {S}t\'efan van~der {W}alt and {J}arrod {M}illman, editors, {\em
  {P}roceedings of the 9th {P}ython in {S}cience {C}onference}, pages 56 -- 61.

\end{thebibliography}
\bibliographystyle{apalike}
\newpage

\appendix
\appendixname{ A: Material Requirements for OSIRIS-REx cleanroom}

\begin{table}[h!]
    \centering
    
    \begin{tabular}{|c|c|}
       \hline 
      \multicolumn{2}{|c|}{\textbf{Materials allowed in OSIRIS-REx glovebox}} \\ \hline \textbf{Category}   & \textbf{Permitted Material} \\\hline
       Glass  & \makecell{Schott Amiran Low Iron Laminated (glovebox windows) \\ Pilkington OpticWhite (uncoated glovebox windows)\\Fused Quartz/Silica\\Borosilicate Crown Glass (BK7)\\Soda-Lime silicate glass\\ Sapphire/Sapphire Glass} \\ \hline Plastics & \makecell{PTFE \\Flourinated Ethylene Propylene \\ Polyvinylidene Flouride \\ Perflouroalkoxy Alkanes} \\ \hline Metals & \makecell{6061, 6063 Aluminum \\ 316, 316l Stainless Steel \\ 304 Stainless Steel \\ 2024 Aluminum \\ 301, 302, 303 Stainless Steel} \\ \hline Metal Finishes & \makecell{Clear and hard Adonized \\ Pickled and Passivated \\ Electro-polished \\ Gold plated (electrical contacts) \\ Satin/Pebble/Media blast} \\ \hline Elastomers & \makecell{Viton (FKM) \\ Chlorosulfonated Polyethylene (CSM, Hypalon)} \\ \hline \multicolumn{2}{|c|}{\textbf{Materials allowed in OSIRIS-REx Cleanroom and glovebox}} \\ \hline \multicolumn{2}{|c|}{Cleanroom Polyester (gowning and similar uses)} \\  \multicolumn{2}{|c|}{PTFE and Kapton Tape} \\  \multicolumn{2}{|c|}{Nitrile gloves} \\ \multicolumn{2}{|c|}{Polycarbonate (Lexan)} \\  \multicolumn{2}{|c|}{Additional stainless-steel alloys} \\  \multicolumn{2}{|c|}{Computers, cameras, and microscopes} \\ \multicolumn{2}{|c|}{LED light sources} \\ \hline      
    \end{tabular}
    \caption{Full list of materials permitted in the ISO-5 OSIRIS-REx cleanroom and the glovebox.}
    \label{tab:materials}
\end{table}

\appendixname{ B: Uncertainty}
\begin{table}
    \centering
    \begin{tabular}{|c|c|}
       \hline \textbf{Source} & \textbf{Uncertainty} \\ \hline
        Detector Nonlinearity & 1\% \\\hline
        Reflectance Standard Calibration & 5\%\\ \hline
        SNR& 2\% \\ \hline
        Flat field efficacy & 2\%  \\ \hline
        Detector Noise & $\leq$ 1\%  \\ \hline
        Sapphire Transmission & 5\%  \\ \hline
    \end{tabular}
    \caption{Uncertainty sources and their contributions to overall uncertainty.}   
    
    \label{tab: err}
\end{table}

The total radiometric uncertainty for QRIS data arises from a variety of sources, summarized in Table \ref{tab: err}.  This analysis was performed before PE and does not include stray light effects that we could not quantify. Detector nonlinearity is corrected within 1\% by our pipeline, and therefore contributes 1\% uncertainty. The 5\% error on reflectance standard calibration arises from the uncertainty in the calibration data provided by Labsphere. After our flat fields are applied to an image of a flat white surface, some regions of the image still deviate from flatness by about 2\%, creating a 2\% uncertainty in flat field efficacy. We assign a 5\% uncertainty to the sapphire transmission values, which are extrapolated from data for similar windows provided by Edmund Optics and verified by our laboratory measurements. 

Detector noise combines bias noise and dark current. As previously explained, dark current is negligible as the camera corrects dark current on-chip.  Bias noise is random electronic noise generated by the detector, which has values of 1-5 DN. We therefore state that detector noise uncertainty is $\leq$1\%.

The Signal to Noise Ratio (SNR) is defined as the ratio between signal and shot noise. We define shot noise as $\sqrt{N}$, where $N$ is the number of photons hitting the detector. Therefore, 
\begin{equation}
    SNR = \frac{N}{\sqrt{N}}.
\end{equation}
Dividing the SNR by the total signal yields the SNR uncertainty. For all bands except UV, the lowest signal that is included in the final HDR image is about 2000 DN, which is the average signal on the 2\% standard in 10 second exposure images. For UV, the average signal on the 2\% standard in 10 second exposure images is about 700 DN. This discrepancy is a result in the camera's lower sensitivity to short wavelengths. For UV, the approximate upper limit on the SNR uncertainty is
\begin{equation}
    \frac{\sqrt{700 \textrm{ DN}}}{700 \textrm{ DN}} = 0.0378 = 3.78 \%.
    \label{uvcalc}
\end{equation}
For all other bands, the approximate upper limit of SNR uncertainty is 
\begin{equation}
    \frac{\sqrt{2000 \textrm{ DN}}}{2000 \textrm{ DN}} = 0.0224 = 2.24 \%.
    \label{allbandscalc}
\end{equation}

Since the values calculated in Equations \ref{uvcalc} and \ref{allbandscalc} are upper limits that only apply to the very darkest parts of the images, we conclude that the average SNR uncertainty is 2\%.

To determine QRIS's total radiometric uncertainty, we use the values from Table \ref{tab: err} to calculate RSS uncertainty. Equation \ref{rss}, where U$_{rad}$ is total radiometric uncertainty, shows this calculation. 
\begin{align}
    \label{rss}
    U_{rad} &= \sqrt{(1)^2 + (5)^2 + (2)^2 + (2)^2 + (1)^2 + (5)^2}, \\
    U_{rad} &= 7.7 \%. \notag
\end{align}
To account for unknown sources of error, we conclude that total radiometric uncertainty for QRIS is $\leq$ 10 \%.

\end{document}